\documentclass[namedreferences]{solarphysics}

\usepackage[hyperref,solaromanenum]{my-sola-addons} 
\usepackage{graphicx}                    
\usepackage{color}                       
\usepackage{breakurl}                         

\usepackage{setspace}

\newcommand{\aap}{    {\it Astron. Astrophys.}}
\newcommand{\aaps}{   {\it Astron. Astrophys. Suppl.}}

\newcommand{\apj}{    {\it Astrophys. J.}}
\newcommand{\apjs}{\aaps}
\newcommand{\apjl}{   {\it Astrophys. J. Lett.}}

\newcommand{\solphys}{{\it Solar Phys.}}

\newcommand{\zap}{\it Zeitschrift Astrophysik}
\chardef\us=`\_

\newenvironment{coi}[1][Disclosure of Potential Conflicts of Interest]{\footnotesize\paragraph*{#1}}{}

\begin{document}

\begin{article}

\begin{opening}

\title{A Fixed-point Scheme for the Numerical Construction
of Magnetohydrostatic Atmospheres in Three
Dimensions}

%
\author[addressref={1},
corref,email={sgilchrist@nwra.com}]
{\inits{S.A.}\fnm{S. A. } \lnm{ Gilchrist}}

\author[addressref={1},
corref,email={}]
{\inits{D.C.}\fnm{D. C. } \lnm{ Braun}}

\author[addressref={1},
corref,email={}]
{\inits{G.}\fnm{G. } \lnm{ Barnes}}

%
\runningauthor{S.A. Gilchrist, D.C. Braun, G. Barnes}
\runningtitle{A Fixed-point Scheme for the Numerical Construction
              of Magnetohydrostatic Atmospheres in Three
              Dimensions}

\address[id={1}]{NorthWest Research Associates (NWRA), 
                 3380 Mitchell Ln., Boulder, CO 80301, 
                 USA}

\begin{abstract}
  Magnetohydrostatic models of the solar atmosphere are 
  often based on idealized analytic solutions because the underlying
  equations are too difficult to solve in full generality. Numerical
  approaches, too, are often limited in scope and have tended to focus on the 
  two-dimensional problem. In this article we develop a numerical 
  method for solving the nonlinear magnetohydrostatic equations
  in three dimensions. Our method is a fixed-point iteration scheme
  that extends the method of Grad and Rubin ({\it Proc. 2nd Int. Conf. on 
  Peaceful Uses of Atomic Energy} {\bf 31}, 190, 1958) to include 
  a finite gravity force. We apply the method to a test case to 
  demonstrate the method in general and our implementation in code 
  in particular. 
\end{abstract}

%

\end{opening}

%

\newcommand{\iumax}{$-1.35$}
\newcommand{\iuavg}{$-1.55$}

%
%

\section{Introduction}

\noindent There is a general interest in constructing magnetohydrostatic
models of the solar atmosphere. These models describe large-scale, long-lived,
magnetic structures like sunspots ({\it e.g.} \citealp{1958IAUS....6..263S}), 
prominences ({\it e.g.} \citealp{1957ZA.....43...36K}), coronal loops 
({\it e.g.} \citealp{1982SoPh...76..261Z}), the  coronal magnetic field on global
scales ({\it e.g.} \citealp{1986ApJ...306..271B,2008A&A...481..827R}),
and low-lying magnetic structures in the upper photosphere and lower
chromosphere ({\it e.g.} \citealp{2015ApJ...815...10W}).
Unfortunately, the equations of the model --- the magnetohydrostatic equations ---
are a set of nonlinear partial differential equations that defy general
analytic solution. Only a handful of idealized analytic solutions are known, and existing
numerical methods are typically limited to two dimensions. Solution
methods for the general three-dimensional equations are lacking, 
which limits the scope of the modeling. In this article, 
we present a new numerical scheme for treating the general 
three-dimensional problem. Our method is an extension of the 
Grad-Rubin method \citep{24756} to include a finite gravity force.\\

\noindent The magnetohydrostatic equations describe a magnetized plasma
in which magnetic, pressure, and external forces are in mechanical 
equilibrium. In the solar context, the external force is usually gravity, 
implying that the condition for mechanical equilibrium is 
\begin{equation}
  \mathbfit J \times \mathbfit B - \nabla p + \rho \mathbfit g = \mathbfit 0
\end{equation}
\citep{9559}, where $\mathbfit J$ is the electric current density, $\mathbfit B$ is
the magnetic field, $p$ is the gas pressure, $\rho$ is the gas density,
and $\mathbfit g$ is the local acceleration due to gravity. 
The special case with $\mathbfit g = \mathbfit 0$ is of interest to modeling fusion plasmas 
({\it e.g.} \citealp{4317418,CHODURA198168,greenpaper}), but in this article 
we consider only the case with a finite gravity force, which is more relevant
to modeling the Sun. Since $\mathbfit J$ is the curl of $\mathbfit B$ in accordance
with Amp{\'e}re's law, the force-balance equation is nonlinear 
through the Lorentz force term. It is this nonlinearity that 
makes the magnetohydrostatic equations difficult to solve.\\

\noindent Magnetohydrostatic models find various applications
in solar physics ({\it e.g.} \citealp{1958IAUS....6..263S,1957ZA.....43...36K,1982SoPh...76..261Z,
1986ApJ...306..271B,2008A&A...481..827R,1986A&A...170..126S,
1971SoPh...18..258P}). One area where they are becoming increasingly
relevant is local helioseismology, where magnetohydrostatic sunspot
models are used as the background atmosphere to 
magnetohydrodynamic wave propagation simulations ({\it e.g.} \citealp{2006ApJ...653..739K,2008SoPh..251..589K,
2009ApJ...690L..72M,2011SoPh..268..293C}). The present
modeling is based on axisymmetric magnetohydrostatic solutions
(see \citealp{2010SoPh..267....1M} for a full list of models) 
and so is limited to monopolar sunspots. To construct atmospheres
for more complex sunspot groups requires more general solution methods. \\

\noindent Data-driven modeling of the coronal magnetic field is
another area where three-dimensional magnetohydrostatic models
have potential applications. The coronal magnetic field is often ``extrapolated'' 
from vector-magnetogram data based on a nonlinear force-free model
({\it e.g.} \citealp{2002A&A...392.1119R,2008A&A...488L..71T,
2012SoPh..276..133G,2012SoPh..278...73V}).
However, it is known that the magnetogram data represent the magnetic
field at a height in the atmosphere where pressure and gravity forces,
which the force-free model excludes, are significant \citep{1995ApJ...439..474M,
2001SoPh..203...71G,2005ApJ...631L.167S}.
This inconsistency is a potential source of problems for the modeling 
(\citealp{2009ApJ...696.1780D},\citeyear{2015ApJ...811..107D}), and a self-consistent 
approach based on a magnetohydrostatic model has been suggested
as a solution but never applied to actual data \citep{2003SoPh..214..287W,2009ApJ...696.1780D}. 
To the best of our knowledge, the only magnetohydrostatic extrapolations
performed to date have been based on a special class of 
``linear'' magnetohydrostatic solution that assumes
a particular functional form for the current density 
\citep{1992ApJ...399..300L,2000A&A...356..735P}.
These solutions have been used to extrapolate the coronal magnetic
field from magnetogram data for several studies
({\it e.g.} \citealp{1998SoPh..183..369A,1999A&A...342..867A,2000PhDT.........2P,2015ApJ...815...10W})\\

\noindent The nonlinearity of the magnetohydrostatic equations 
complicates their solution, and no method is known for constructing
general analytic solutions. Particular analytic solutions, however, can be 
derived by simplifying the equations at the expense
of generality. One strategy 
is to reduce the dimensionality of the problem by assuming self-similarity
\citep{1958IAUS....6..263S,1980SoPh...67...57L}, 
translational symmetry \citep{1982SoPh...76..261Z}, 
or rotational symmetry \citep{1981JApA....2..405U}.
Another approach, which produces three-dimensional solutions, is to 
impose a special form on either the current density \citep{1985ApJ...293...31L,
1997A&A...325..847N,2000A&A...356..735P} 
or the magnetic tension \citep{1984ApJ...277..415L}. All these solutions
are special cases --- the general three-dimensional
magnetohydrostatic problem remains unsolved.\\

\noindent A numerical approach can in principle address the
shortcomings of the analytic methods. With this goal, a number
of numerical methods have been developed, although a majority have only 
been implemented in two dimensions. These methods include 
magnetohydrodynamic relaxation methods ({\it e.g.} \citealp{1984A&A...139..426D}), 
fixed-point iteration methods ({\it e.g.} 
\citealp{1986ApJ...302..785P,1990ApJ...365..764P}), and 
nonlinear multigrid methods applied to the 
magnetohydrostatic equations formulated in inverse coordinates
({\it e.g.} \citealp{1990JCoPh..89..490C}). This list is not exhaustive:
\citet{2001SoPh..201..289H} provide a more complete list 
of references.\\

\noindent Less work has been done on numerical methods for the 
three-dimensional problem. \citet{2001SoPh..201..289H} develop
a three-dimensional inverse coordinate nonlinear multigrid method.
Their method also solves the free-surface problem for a flux rope
bounded by a current sheet. The optimization method that was 
originally introduced by \citet{2000ApJ...540.1150W} for solving
the nonlinear force-free equations has been extended to treat 
the magnetohydrostatic equations in Cartesian \citep{2003SoPh..214..287W}
and spherical coordinates \citep{2007A&A...475..701W}. \\

\noindent In this article we develop a fixed-point method for solving
the general three-dimensional magnetohydrostatic equations in a 
Cartesian box. In particular, we extend the method of 
\citet{24756} to model a gravity force.  The original Grad-Rubin method solves
the magnetohydrostatic equations without gravity by 
replacing the nonlinear equations by a system of linear equations
for each unknown variable, which are solved interactively. 
The linearization is achieved by constructing nonlinear terms 
using variables from previous iterations. For example, the Lorentz
force is constructed as 
\begin{equation}
  \nabla \times \mathbfit B^{[k+1]} \times \mathbfit B^{[k]},
\end{equation}
which is linear in $\mathbfit B^{[k+1]}$, since $\mathbfit B^{[k]}$
is known from a previous iteration (the superscript denotes the 
iteration number). This makes it possible to
define a system of linear ``update equations'' that relate each
variable at the current iteration to those known from previous iterations.
The update equations are solved successively, and a solution to 
the complete nonlinear system is obtained when and if the iteration
converges to a fixed point. This method has been previously used
to solve the magnetohydrostatic equations with $\mathbfit g = 0$
\citep{greenpaper,doi:10.1137/070700942,2013SoPh..282..283G}.
It has also been used to solve the nonlinear force-free equations
\citep{1981SoPh...69..343S,1999A&A...350.1051A,2004SoPh..222..247W,
2007SoPh..245..251W}, which is the special case of the magnetohydrostatic
equations defined by the condition $\mathbfit J \times \mathbfit B = 0$.
In this article we extend the method to solve
the magnetohydrostatic equations with $\mathbfit g \ne 0$. \\

\noindent Constructing solutions using a fixed-point method has 
several potential advantages over the methods of 
\citet{2001SoPh..201..289H} and \citet{2003SoPh..214..287W}. 
Firstly, unlike methods formulated in inverse coordinates, 
our method does not fix the toplogy of the magnetic field {\it a priori}.
Secondly, the method does not overspecify the boundary-value 
problem like the optimization method.\\

%
%

\section{The Magnetohydrostatic Model and Boundary-Value Problem}
\label{sec_equations_and_model}

\noindent In this section we describe the details of the 
magnetohydrostatic model. We present the equations and formulate
the magnetohydrostatic boundary-value problem. \\

\noindent The magnetohydrostatic equations in Cartesian coordinates
are \citep{9559}:
\begin{equation}
  \mathbfit J \times \mathbfit B - \nabla p - \rho g(z) \hat{\mathbfit{z}} = 0,
  \label{mhs_eq1}
\end{equation}
\begin{equation}
  \nabla \times \mathbfit B = \mu_0 \mathbfit J,
  \label{mhs_eq2}
\end{equation}
and 
\begin{equation}
  \nabla \cdot \mathbfit B = 0,
  \label{mhs_eq3}
\end{equation}
where $\mathbfit B$ is the magnetic field, $\mathbfit J$ is the 
electric current density, $p$ is the gas pressure, $\rho$ is
the gas density, and $g(z)$ is the acceleration due to 
gravity. For our model we will assume that the force of gravity acts
in the negative $z$ direction and is a known function of $z$. \\ 

\noindent In addition to Equations (\ref{mhs_eq1})--(\ref{mhs_eq3}),
it is also necessary to specify an equation of state and an energy equation
to close the system. For the equation of state, we use the ideal gas law,
\begin{equation}
  p(\mathbfit r) = \frac{R}{\tilde\mu(\mathbfit r)} \rho (\mathbfit r) T(\mathbfit r),
\end{equation}
where $R$ is the ideal gas constant, and $\tilde\mu(\mathbfit r)$ is the
mean atomic weight, which may vary with position $\mathbfit r$.
Since only the ratio
$T/{\tilde \mu}$ is important, we introduce the scale height,
\begin{equation}
  h(\mathbfit r) = \frac{R T(\mathbfit r)}{\tilde{\mu}(\mathbfit r) g(z)},
\end{equation}
and rewrite the equation of state as
\begin{equation}
  p(\mathbfit r) = g(z) \rho(\mathbfit r) h(\mathbfit r).
\end{equation}

\noindent The final equation is the steady-state energy equation.
A realistic equation, however, is nontrivial to construct.
In the interior, energy transport by convection is important, but is difficult 
to treat self-consistently without modeling flows. 
Higher in the atmosphere, coronal heating must be treated, which 
is difficult because the underlying mechanism is uncertain.
Although there are some exceptions 
\citep{1993ApJ...404..788P,1990SoPh..126...69F}, explicit treatment
of the energetics is often avoided for these reasons.
For example, the method of \citet{2007A&A...475..701W} 
models neither an energy equation nor an equation of state, and the 
method of \citet{1986ApJ...302..785P} treats energy transport 
implicitly by prescribing $h(\mathbfit r)$ in the volume. For our model 
we adopt the latter approach and prescribe $h(\mathbfit r)$ everywhere in the 
computational volume.\\

\noindent In summary, the dependent variables of the model are 
$\mathbfit B$, $\mathbfit J$, $p$ and $\rho$. It is assumed that 
both the scale height $h(\mathbfit r)$ and the acceleration due to gravity 
$g(z)$ are known everywhere in the volume, and so given either 
$p$ or $\rho$, the other is known.\\

\subsection{Domain and Boundary-Value Problem}
\label{sec_bvp}

\noindent We solve the model in the finite Cartesian box
\begin{equation}
  V = \left \{ (x,y,z) \left |  
                         -\frac{L_x}{2} \le x \le +\frac{L_x}{2}, 
                         -\frac{L_y}{2} \le y \le +\frac{L_y}{2},
                                0 \le z \le +L_z
      \right.
      \right \}.
  \label{eq_domain}
\end{equation}
The boundary of the domain, $\partial V$, is formed by the six
planar faces of the box. \\
Working in Cartesian coordinates simplifies the numerical 
implementation, but this geometry is unsuitable for modeling large-scale
structures or the global magnetic field because the curvature of
the Sun cannot be ignored at these scales. A correct treatment requires an implementation
in spherical geometry, such as the magnetohydrostatic method
of \citet{2007A&A...475..701W} or the force-free methods 
of either \citet{2013A&A...553A..43A} or \citet{2014SoPh..289.1153G}.
Nevertheless, the Cartesian implementation can still be useful
for structures of active-region size.\\

\noindent To solve the magnetohydrostatic equations, boundary
conditions are required on a subset of the dependent variables 
$\mathbfit B$, $\mathbfit J$, $p$, and $\rho$. Since the pressure
and density are related through the ideal gas law and since 
$h(\mathbfit r)$ is prescribed everywhere, $p$ and $\rho$ are not independent --- 
fixing both independently at the boundary is an overspecification.
Of the two, we prescribe boundary conditions on $p$.\\

\noindent All that remains is to specify the exact form of the 
boundary conditions on $\mathbfit B$, $\mathbfit J$ and $p$. We follow
the approach of \citet{24756}, who consider different formulations
of the boundary-value problem for the special case with $\mathbfit g=0$.
The Grad-Rubin boundary conditions are the normal component of $\mathbfit B$,
\begin{equation}
  \left. \mathbfit B \cdot \hat{\mathbfit{n}} \right |_{\partial V} = B_n,
  \label{bcs_bn}
\end{equation}
the normal component of $\mathbfit J$,
\begin{equation}
  \left. \mathbfit J \cdot \hat{\mathbfit{n}} \right |_{\pm \partial V} = J_n, 
\end{equation}
and the pressure distribution,
\begin{equation}
  \left. p \right |_{\pm \partial V} = p_0.
\end{equation}
Here $\pm \partial V$ is either $+\partial V$ or $-\partial V$, which
are the subsets of $\partial V$ where $B_n > 0$ and where $B_n < 0$ respectively.
This means that the boundary conditions on the pressure and current
density are prescribed either at points in the boundary where $B_n>0$ 
or at points in the boundary where $B_n<0$, but not both. The reason 
for this is that pressure and the field-aligned component of the 
current density obey hyperbolic transport equations along field lines
(see \citealp{24756}), and thus boundary conditions on  $J_n$ and $p$  
must only be prescribed at one end of each field line to avoid
overspecification of the boundary-value problem.\\

\noindent In formulating the boundary-value problem, we also make
the assumption that all field lines are connected to the boundary.
This is again related to the hyperbolic character of the underlying
equations --- information in the boundary is transported into the
volume along magnetic field lines, meaning points not connected to 
the boundary are undetermined. Formulations that account for closed
field lines are possible and are discussed by \citet{24756}.
Since closed field lines are not considered, imposing boundary conditions at points where $B_n=0$ is also 
an overspecification, as field lines threading these points cross 
the boundary elsewhere, and therefore prescribing boundary
conditions where $B_n=0$ violates the requirement that there be
only one set of boundary conditions {\it per} field line.\\

\noindent Here we have presented a basic formulation of the boundary-value 
problem, however more complicated ``self-consistency'' methods that involve
the solution of a sequence of boundary-value problems have been
developed for the force-free equations \citep{2009ApJ...700L..88W}.
We note that a similar approach could be developed for the 
magnetohydrostatic problem. 


\subsection{Pressure and Density Decomposition}
\label{sec_pd_decomp}

\noindent Rather than working with $p$ and $\rho$ directly, it is 
advantageous to reformulate the magnetohydrostatic equations in
terms of the deviation from a gravitationally stratified 
background atmosphere. The pressure and density deviation due
to the magnetic field can be orders of magnitude smaller than the pressure and
density of the background and therefore are difficult to resolve numerically.
In fact, the difference in magnitudes can lead to numerical instability, as 
discussed in Appendix A. Splitting $p$ and $\rho$ into background
and magnetohydrostatic components and computing each separately 
mitigates this problem.\\

\noindent The magnetohydrostatic equations are linear in 
$p$ and $\rho$, so we may write
\begin{equation}
  p(\mathbfit r) = p_{\rm hs}(z) + p_{\rm mhs}(\mathbfit r),
\end{equation}
and 
\begin{equation}
  \rho(\mathbfit r) = \rho_{\rm hs}(z) + \rho_{\rm mhs}(\mathbfit r),
\end{equation}
where $p_{\rm hs}(z)$ and $\rho_{\rm hs}(z)$ are due to a background 
atmosphere and $p_{\rm mhs}$ and $\rho_{\rm mhs}$
are the variations due to the presence of the magnetic field.
Note that the magnetohydrostatic components can be negative.\\

\noindent The background pressure and density satisfy the hydrostatic 
force-balance equation 
\begin{equation}
  \frac{dp_{\rm hs}}{dz} = - g(z)\rho_{\rm hs}(z),
  \label{fb_hs}
\end{equation}
and the equation of state 
\begin{equation}
  p_{\rm hs}(z) = g(z)\rho_{\rm hs}(z) h_{\rm hs}(z)
  \label{eos_hs}
\end{equation}
\noindent where $h_{\rm hs}(z)$ is the scale height for the
background atmosphere. To uniquely determine $p_{\rm hs}$ and $\rho_{\rm hs}$,
it is necessary to prescribe the scale height  $h_{\rm hs}(z)$
in the volume and the pressure at either the top or bottom boundary, {\it i.e.}
\begin{equation}
  \left. p_{\rm hs} \right |_{z=0/L_z} = p_{0 \rm hs}.
\end{equation}
From a purely mathematical standpoint, $h_{\rm hs}(z)$ 
and $p_{0 \rm hs}$ may be chosen with some freedom. However, for 
numerical reasons, we choose to construct $p_{\rm hs}$ and $\rho_{\rm hs}$ as the pressure 
and density of the ``quiet Sun''
and  $p_{\rm mhs}$ and $\rho_{\rm mhs}$ as deviations from this 
background. In this context ``quiet Sun'' refers to regions where
the magnetic field is negligible. This interpretation of $p_{\rm hs}$
and $\rho_{\rm hs}$ requires that $h_{\rm hs}$ be
the asymptotic form of $h(\mathbfit r)$ in regions where
the magnetic field is small. With this choice, $p_{\rm mhs}$ and $\rho_{\rm mhs}$
are generally small in weak-field regions, which turns out to be important
numerically (see Appendix A).\\

\noindent With the pressure and density split, the magnetohydrostatic
force-balance equation becomes
\begin{equation}
  \mathbfit J \times \mathbfit B - 
  \nabla p_{\rm mhs} - 
  \rho_{\rm mhs} g(z) \hat{\mathbfit {z}} = 0,
\end{equation}
and the ideal gas law becomes
\begin{equation}
  p_{\rm mhs} = g(z)h(\mathbfit r) \rho_{\rm mhs}(\mathbfit r)
              + g(z)[h_{\rm hs}(z)- h(\mathbfit r)] \rho_{\rm hs}.
\end{equation}
The boundary conditions for $p_{\rm mhs}$ are then
\begin{equation}
  \left. p_{\rm mhs} \right |_{\pm \partial V} = 
   p_0 - \left. p_{\rm hs} \right |_{\pm \partial V}.
\end{equation}
We use the split form of the magnetohydrostatic equations to
formulate our fixed-point iteration.

%
%

\section{Iteration Scheme}
\label{sec_iter_method}

\noindent In this section we describe the fixed-point iteration 
scheme for solving the boundary value problem presented in 
Section \ref{sec_bvp}.  The numerical implementation of the 
method is presented later in Section \ref{sec_numerical}.\\ 

\noindent To solve the model presented in Section \ref{sec_equations_and_model},
we employ an iterative scheme that extends the method 
of \citet{24756} to include a gravity force.
The method replaces the system of nonlinear equations
with a set of linear equations that are easier to solve than the 
original nonlinear ones. The linear equations are solved 
iteratively and the solution to the nonlinear system is obtained
when (and if) the iteration reaches a fixed point.\\

\noindent In the following we describe the steps in a single iteration
of the method. We denote a variable after $k$ iterations by a superscript 
in square brackets, {\it e.g.} $\mathbfit B^{[k]}$ denotes the magnetic field 
after $k$ iterations. In addition, we describe the magnetic field
used to initialize the method.\\


\subsection{Initial Magnetic Field}
\label{sec_pot_field}

\noindent In principle the iteration can be initiated with
any magnetic field that satisfies the boundary conditions
given by Equation (\ref{bcs_bn}), but in practice the 
simplest field to use is a current-free 
(potential) magnetic field $\mathbfit B^{[0]} = \mathbfit B_0$. 
This choice may be suboptimal, as other choices, like a nonlinear
force-free field, may lie closer to the fixed-point and therefore
reduce the number of iterations required to reach a solution.
The advantage of using a potential field is that it is straightforward
and fast to compute.\\

\noindent The potential field is found  by solving the magnetostatic
equations \citep{9559}
\begin{equation}
  \nabla \cdot \mathbfit B_0 = 0
  \label{eq_pot1}
\end{equation}
and
\begin{equation}
  \nabla \times \mathbfit B_0 = 0,
  \label{eq_pot2}
\end{equation}
subject to the boundary conditions 
\begin{equation}
  \left . \mathbfit B_0 \cdot \hat{\mathbfit{n}} \right |_{\partial V} = B_n.
\end{equation}
Since the normal component of the magnetic field is prescribed on
all six boundaries, it is necessary that the net flux over all the 
boundaries be zero in order for $\mathbfit B_0$ to be solenoidal in the volume. \\


\subsection{Iteration Steps}
\label{sec_iter_basic}

\noindent A single iteration of the scheme involves five steps.
At each step the new value of a variable is computed by solving
either an algebraic or differential equation. It is assumed that
at the start of each iteration the magnetic field from the previous
iteration is known. The potential field described in Section
\ref{sec_pot_field} is used for the first iteration. In addition,
it is assumed that $\rho_{\rm hs}$ and $p_{\rm hs}$ have been 
computed beforehand.\\

\noindent Here we enumerate the steps for a single iteration with
iteration number $k$. The updated variables have iteration number 
$k+1$. The steps are as follows.\\

\begin{enumerate}


\item \label{p_update} Calculate $p_{\rm mhs}^{[k+1]}$, the pressure in 
the volume, by solving 
\begin{equation}
  \nabla p^{[k+1]}_{\rm mhs} \cdot \mathbfit B^{[k]} = 
  \left (\frac{B_z^{[k]}}{h(\mathbfit r)} \right ) p^{[k+1]}_{\rm mhs}
  + \left ( \frac{p_{\rm hs}}{h(\mathbfit r)} - \rho_{\rm hs}(z)g(z) \right ) 
           B_z^{[k]},
  \label{iter_eq1}
\end{equation}
subject to the boundary conditions for $p_{\rm mhs}$ described in 
Section \ref{sec_bvp}. Since $\rho_{\rm hs}$, $p_{\rm hs}$,
$h(\mathbfit r)$, $g(z)$, and $B_z^{[k]}$ are known, Equation (\ref{iter_eq1}) is 
linear.

\item \label{rho_update} Calculate $\rho^{[k+1]}_{\rm mhs}$, the 
gas density in the volume, by direct application of the ideal gas law
\begin{equation}
  \rho_{\rm mhs}^{[k+1]} = \frac{p_{\rm mhs}^{[k+1]}}{g(z)h(\mathbfit r)}
                 + \frac{p_{\rm hs}}{g(z)}\left ( \frac{1}{h(\mathbfit r)}
                 - \frac{1}{h_{\rm hs}(z)} \right ).                 
  \label{iter_eq2}
\end{equation}


\item \label{jp_update} Calculate $\mathbfit J_{\perp}^{[k+1]}$,
the component of current density that is perpendicular 
to the magnetic field, in the volume. This is computed as
\begin{equation}
   \mathbfit J_{\perp}^{[k+1]} = \frac{
                             - \nabla p^{[k+1]}_{\rm mhs} \times \mathbfit B^{[k]}
                             + \rho^{[k+1]}_{\rm mhs} \mathbfit g \times \mathbfit B^{[k]}
                            }{ 
                            ||\mathbfit B^{[k]}||^2 
                            }.
  \label{iter_eq3}                            
\end{equation}


\item \label{a_update} Calculate $\mathbfit J_{\parallel}^{[k+1]}$,
the component of the current density that is parallel to the magnetic field,
in the volume. This component can be expressed as 
\begin{equation}
  \mathbfit J_{\parallel}^{[k+1]} = \sigma^{[k+1]}\mathbfit B^{[k]},
\end{equation}
where $\sigma(\mathbfit r)^{[k+1]}$ is a scalar function that varies
with position and satisfies the hyperbolic equation
\begin{equation}
  \nabla \sigma^{[k+1]} \cdot \mathbfit B^{[k]} = - \nabla \cdot \mathbfit J_{\perp}^{[k+1]}.
  \label{iter_eq_alpha}
\end{equation}
The component $\mathbfit J_{\parallel}^{[k+1]}$ is found by solving 
Equation (\ref{iter_eq_alpha}) subject to boundary conditions
derived from those on $J_n$ and $B_n$, {\it i.e.}
\begin{equation}
  \left .  \sigma^{[k+1]} \right |_{\pm \partial V} = 
  \left. \frac{\mathbfit J_{\perp}^{[k+1]} \cdot 
  \hat{\mathbfit{n}} - J_n}{B_n} \right |_{\pm \partial V}.
\end{equation}


\item \label{b_update} Calculate $\mathbfit B^{[k+1]}$, the new magnetic field in the
volume, by solving Amp\`ere's law, 
\begin{equation}
  \nabla \times \mathbfit B^{[k+1]} = \mu_0 \mathbfit J^{[k+1]},
  \label{iter_eq_ampere}
\end{equation}
subject to the solenoidal condition
\begin{equation}
  \nabla \cdot \mathbfit B^{[k+1]} = 0,
\end{equation}
and the boundary conditions on the normal component described
in Section \ref{sec_bvp}. The current density in Equation (\ref{iter_eq_ampere}) 
is constructed from the components calculated in steps \ref{jp_update})
and \ref{a_update}), {\it i.e.}
\begin{equation}
  \mathbfit  J^{[k+1]} = \mathbfit J_{\perp}^{[k+1]}
                     + \sigma^{[k+1]} \mathbfit B^{[k+1]}/\mu_0.
\end{equation}

\end{enumerate}

%
%

\section{Numerical Implementation}
\label{sec_numerical}

\noindent This section describes the implemenation of 
the scheme presented in Section \ref{sec_iter_method} in code.\\

\noindent The code is written in a combination of the Fortran 2008 
\citep{Metcalf:2011:MFE:2090092} and C programming languages.
It is parallelized for shared-memory parallel computers 
using OpenMP \citep{Chandra:2001:PPO:355074}. The code 
supports calculations in either single, double, extended, or quadruple
precision, although use of the latter two slows the calculation.\\

\noindent The numerical mesh is constructed so that the spacing is 
uniform in any particular dimension but may differ between dimensions.
In units of grid points, the domain has the total volume 
$N_x \times N_y \times N_z$. Each mesh point is 
\begin{equation}
  (x_i,y_j,z_k) = (i\Delta_x,j\Delta_y,k \Delta_z),
\end{equation}
where $0 \le i < N_x$, $0 \le j < N_y$, and
$0 \le k < N_z$. The grid spacing in each dimension is 
\begin{equation}
  \Delta_l = \frac{L_l}{N_l-1},
\end{equation}
where $l=x,y,$ or $z$.


\subsection{Solution of Elliptic Equations}

\noindent Calculation of both the potential field and the
non-potential field requires the solution of a set of elliptic
partial differential equations. By introducing the appropriate
scalar and vector potentials, both problems can be reduced to the problem
of solving Poisson's equation. \\

\noindent For the potential field calculation we introduce the 
scalar potential, $\phi_{\rm m}$, defined by  
\begin{equation}
  \mathbfit B_0 = \nabla \phi_{\rm m}.
\end{equation}
This reduces the problem of solving the potential-field boundary-value 
problem from Section \ref{sec_pot_field} 
to solving Laplace's equation
\begin{equation}
  \nabla^2 \phi_{\rm m} = 0, 
\end{equation}
subject to the Neumann boundary conditions, 
\begin{equation}
  \left . \nabla \phi_{\rm m} \cdot \hat{\mathbfit{n}} \right |_{\rm \partial V} = B_n.
\end{equation}

\noindent For the magnetic field calculation in step \ref{b_update}), 
only the non-potential component of the magnetic field needs to be 
updated. For this reason, we express $\mathbfit B^{[k+1]}$ as
\begin{equation}
  \mathbfit B^{[k+1]} = \mathbfit B_0 + \nabla \times \mathbfit A_c^{[k+1]}, 
\end{equation} 
where $\mathbfit B_0$ is the potential field defined in Section 
\ref{sec_pot_field}, and $\mathbfit A_c^{[k+1]}$ is a magnetic
vector potential. In the Coulomb gauge ($\nabla \cdot \mathbfit A_c^{[k+1]}=0$),
$\mathbfit A_c^{[k+1]}$  is a solution of Poisson's equation,
\begin{equation}
  \nabla^2 \mathbfit A^{[k+1]} = -\mu_0 \mathbfit J^{[k+1]}.
\end{equation}
Since $\mathbfit B_0$ is constructed to satisfy the boundary conditions 
defined by Equation (\ref{bcs_bn}), it follows that the normal component
of the non-potential component must vanish on the boundary.
The boundary conditions on $\mathbfit A_c^{[k+1]}$ that achieve this
are:
\begin{equation}
  \left .
  \mathbfit A_c^{[k+1]} \times \hat{\mathbfit{n}}\right |_{\partial V}  = 0,
\end{equation}
and 
\begin{equation}
  \left .
  \hat{\mathbfit{n}} \cdot \nabla (\mathbfit A_c^{[k+1]} \cdot \hat{\mathbfit{n}})
  \right |_{\partial V} = 0.
\end{equation}

\noindent Similar to \citet{1986ApJ...302..785P}, we employ 
under relaxation during the magnetic-field-update step in 
order to improve the numerical stability, {\it i.e.} rather than
update $\mathbfit B^{[k+1]}$ directly, we set
\begin{equation}
  \mathbfit B^{[k+1]} = (1-\omega)\mathbfit B^{[k]}
                    + \omega(\mathbfit B_0 + \nabla \times \mathbfit A_c^{[k+1]}),
\end{equation}
where $\omega$ is a constant in the range $(0,1]$. We find
in practice that this improves stability of the iteration scheme 
without altering the fixed point.\\

\noindent Many numerical methods exist for solving the Laplace
and Poisson equations \citep{Press:2007:NRE:1403886}.
We use a second-order finite difference multigrid scheme \citep{Briggs:2000:MT:357695}.
The method is fast and scales well. For a mesh with uniform
spacing in all dimensions, the time to compute the solution 
has $\sim N^3$ scaling, where $N_x = N_y = N_z = N$.


\subsection{Solution of Hyperbolic Equations}

\noindent  We employ a characteristic (field line tracing) method
to solve the hyperbolic equations. We only present  
the numerical solution of Equation (\ref{iter_eq1}), as
the method for Equation (\ref{iter_eq_alpha}) is similar.
The partial differential equation, Equation (\ref{iter_eq1}), can be recast
as a system of ordinary differential equations along each field line, 
{\it i.e.} the equation for $p^{[k+1]}_{\rm mhs}$ can be expressed as 
\begin{equation}
  \frac{dp_{\rm mhs}}{ds} + F(s)p_{\rm mhs} = G(s), 
  \label{pc_update_ds}
\end{equation}
where 
\begin{equation}
  F(s) = \frac{B_z(s)}{||\mathbfit B(s)|| h(s)}
  \label{eq_hyper_f}
\end{equation}
and 
\begin{equation}
  G(s) =  \left ( \frac{p_{\rm hs}(s)}{h(s)} - \rho_{\rm hs}(s)g(s) \right ) 
          \frac{B_z(s)^{[k]}}{||\mathbfit B(s)^{[k]}||}.
  \label{eq_hyper_g}
\end{equation}
Here $d/ds$ is the derivative along the length of a magnetic 
field line whose path is a solution of the field line equation 
\begin{equation}
  \frac{d \mathbfit x(s)}{ds} = \frac{\mathbfit B^{[k]}}{||\mathbfit B^{[k]}||}, 
  \label{eq_fline}
\end{equation}
where $\mathbfit x(s) = (x(s),y(s),z(s))$ is the Cartesian position
along the field line. \\

\noindent The pressure is updated at each step by solving Equations
(\ref{pc_update_ds}) and (\ref{eq_fline}) for each point in the volume.
For each mesh point $(x_i,y_i,z_i)$ in the volume, we solve 
Equation (\ref{eq_fline}) to determine the path of the field line
that threads $(x_i,y_i,z_i)$ while simultaneously integrating Equation
(\ref{pc_update_ds}) along this path. The calculation is halted when the field 
line crosses the boundary of the domain. We perform the calculation
in both directions along the field lines and only impose boundary
conditions at the endpoint on the boundary with the 
correct polarity for the boundary conditions.\\

\noindent We perform the integration using a 
fourth-order Runge-Kutta integrator with a fixed step size
\citep{Press:2007:NRE:1403886}. Since the path of the field line is 
not restricted to the numerical grid, trilinear interpolation
is used to compute $\mathbfit B$, and $h$ during the integration
\citep{Press:2007:NRE:1403886}. 
We use an ``event-location'' method that combines interpolation
and root finding to determine the location where
the field line crosses the boundary  \citep{Hairer:1993:SOD:153158}.\\

\noindent Solving the hyperbolic system using Runge-Kutta methods
is nontrival because it is not a standard initial-value problem: 
$p^{[k+1]}_{\rm mhs}$ is unknown at the initial point $s=0$.
Instead it is a boundary-value problem with boundary conditions on $p^{[k+1]}_{\rm mhs}$
and $\mathbfit x(s)$ known at different ends of each field line.
At the start of the field line ($s=0$), the value of $\mathbfit x(s)$ is 
known, {\it i.e.} $\mathbfit x(0) = (x_i,y_j,z_k)$, but the value of 
$p^{[k+1]}_{\rm mhs}(0)$ is not, in fact,
this is the value we wish to compute. At the point where
the field line crosses the boundary $(s=s_0$), 
$p^{[k+1]}_{\rm mhs}(s_0)$ is known from the boundary conditions on 
$p_{\rm mhs}$, but the crossing point $\mathbfit x(s_0)$ is a priori
unknown. In this form the boundary-value problem cannot be solved
using standard numerical methods for ordinary-differential equations
({\it e.g.} Runge-Kutta methods). To solve this problem, we introduce
two auxiliary initial-value problems whose solutions, when combined,
give the solution to the boundary-value problem. The auxiliary problems are of the standard
form and can be treated with standard integration methods.
This method is explained in Appendix B.\\


\subsection{Construction of the Background Atmosphere}

\noindent The background atmosphere is constructed by solving 
Equation (\ref{fb_hs}) using numerical quadrature. In order
that $h_{\rm hs}$ approaches $h(\mathbfit r)$ 
in weak-field regions (see discussion in Section \ref{sec_pd_decomp})
we take
\begin{equation}
  h_{\rm hs}(z) = h \left (-\frac{L_x}{2},-\frac{L_y}{2},z \right ),
\end{equation}
where the point $(x,y) = (-L_x/2,-L_y/2)$ is at the edge of the computational 
volume defined by Equation (\ref{eq_domain}). For the boundary 
conditions for Equation (\ref{fb_hs}), we take
\begin{equation}
  \left . p_{\rm hs} \right |_{z = 0} = p_0\left (-\frac{L_x}{2},-\frac{L_y}{2}\right ).
\end{equation}
With this choice it is necessary that regions of strong magnetic
field be isolated from the boundaries. We use this corner point
to define the background field for the calculations presented 
here, but it may not be appropriate in all cases and other choices
are possible.\\

\noindent We have chosen to compute the background atmosphere using
numerical quadrature as this method is applicable  when
$h_{\rm hs}(z)$ is given in table form, which is the 
case for many popular model atmospheres ({\it e.g.} \citealp{1981ApJS...45..635V}).
The code is not restricted to using tabulated background atmospheres 
and can use analytic background atmospheres too. 


\subsection{Other Equations}

\noindent The other steps involve the straightforward 
evaluation of algebraic equations. The exception is 
step \ref{jp_update}), which requires the evaluation of numerical
derivatives. We use a fourth-order
finite-difference method to evaluate derivatives \citep{Press:2007:NRE:1403886}. 
In addition, when computing $\mathbfit J_{\rm \perp}$ we use a ``safety factor''
$\eta$ in the denominator 
\begin{equation}
   \mathbfit J_{\perp}^{[k+1]} = \frac{
                             - \nabla p^{[k+1]}_{\rm mhs} \times \mathbfit B^{[k]}
                             + \rho^{[k+1]}_{\rm mhs} \mathbfit g \times \mathbfit B^{[k]}
                            }{ 
                            ||\mathbfit B^{[k]}||^2  + \eta
                            }.
\end{equation}
This is needed in weak-field regions, where the truncation error 
in the numerical derivative can have a similar effect to that discussed
in Appendix A.

%
%

\section{Application of the Method to an Analytic Test Case}

\noindent We apply our code to a problem with a known analytic 
solution in order to test the fixed-point method in general
and our implementation in code in particular. 
The goal is to show firstly that the scheme converges to a fixed point, 
and secondly that the numerical solution obtained thereby 
is consistent with the known solution within the margins of 
numerical error. In this way we establish that both the method and our code
work. We perform calculations at four different resolutions 
to demonstrate the scaling of the numerical truncation error. We
also determine how the total execution time of the code scales with
grid size.\\

\noindent The calculation is performed in non-dimensional units, but
here we use dimensional quantities expressed in terms of unspecified
characteristic values. For example, the domain size is given as $L_x = 30 L_c$,
where $L_c$ is unspecified.\\


\subsection{Test Case and Calculation Parameters}

\noindent To test the code we use the analytic magnetohydrostatic
sunspot model of \citet{1980SoPh...67...57L}, which
belongs to the class of self-similar solutions 
found by \citet{1958IAUS....6..263S}. In cylindrical polar 
coordinates, the magnetic field for this solution is 
\begin{equation}
  \mathbfit B = 2B_0 
            \exp \left ( -\frac{R^2}{a^2+z^2}                          \right )
                 \left [ \frac{Rz}{(a^2+z^2)^2},0,\frac{1}{a^2+z^2}   \right ]
\end{equation}
\citep{1980SoPh...67...57L}, where $R^2 = x^2 + y^2$, $a$ is a free parameter
that sets the width of the spot, and $B_0$ is a free parameter that 
determines the maximum field strength. The solution is axisymmetric 
and untwisted ($\sigma \equiv 0$).\\

\noindent For the tests, we use a solution with $a=5L_c$, and 
$B_0 = a^2 B_c$, where $L_c$ and $B_c$ are an unspecified characteristic 
length scale and magnetic field strength, respectively. The dimensions of the domain 
are $L_x = L_y = 30L_c$, and $L_z = 5L_c$.  We perform four calculations at different resolutions
with 20 iterations of the method applied starting from a potential
field in each case. The number of grid points in each direction are equal,
{\it i.e.} $N_x = N_y = N_z = N$, but $N$ takes the values 
$64$, $96$, $128$, and $256$ in the four cases.


\subsection{Convergence of the  Fixed-point Scheme}
\label{sec_convergence}

\noindent We first demonstrate that the iteration has converged to a 
fixed point by measuring the change in the magnetic field between
iterations using the metric 
\begin{equation}
  \Delta^{[k]}B_{\rm max} = \mbox{max}(|\mathbfit B^{[k+1]} - \mathbfit B^{[k]}|),
  \label{eq_con_metric}
\end{equation}
where $||$ is the componentwise absolute value not the vector norm,
and $\mbox{max}()$ is the maximum value over the domain and three components.
The value of $\Delta^{[k]}B_{\rm max}$ is an upper bound on the 
pointwise change and is therefore small only when the change at 
every point is small. We regard an iteration as converged when $\Delta^{[k]}B_{\rm max}$ 
becomes small. This is a strict measure of convergence,
as $\Delta^{[k]}B_{\rm max}$ will not decrease unless all three
components of $\mathbfit B^{[k]}$ converge to a solution at every 
point in the volume.\\

\begin{figure} 
  \centerline{\includegraphics[width=1.\textwidth]{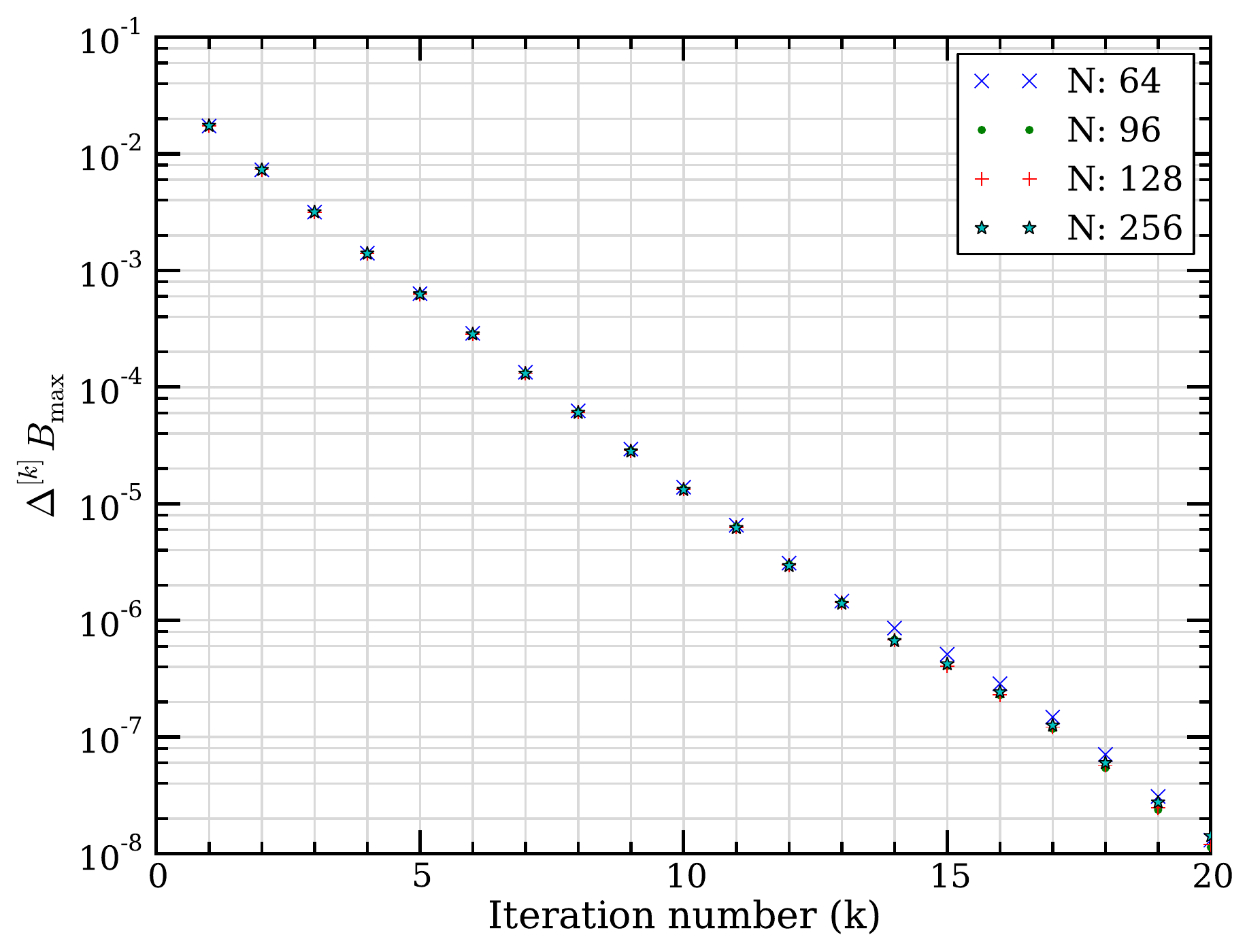}}
  \caption{Change in the magnetic field between successive iterations
           as measured by the $\Delta B_{\rm max}$ metric for the
           four test cases. Each calculation is performed with a 
           different grid size $N$.}                      
  \label{fig_conv}
\end{figure}

\noindent Figure \ref{fig_conv} shows $\Delta^{[k]} B_{\rm max}$ 
{\it versus} iteration number $k$ for the four test cases. In each case, 
$\Delta^{[k]} B_{\rm max}$ decreases monotonically by
about six orders of magnitude over 20 iterations. Given
the small value of $\Delta^{[k]} B_{\rm max}$, we regard the iteration
as converged in all four cases.


\subsection{Measurements of the Numerical Truncation Error}

\noindent We next measure the 
numerical truncation error by comparing the numerical solution 
at iteration 20 to the known analytic solution. To measure 
the difference between the analytic and numerical solutions, we use
the two metrics
\begin{equation}
  u_{\rm max} = \mbox{max}(|\mathbfit B - \mathbfit b |)
\end{equation}
and
\begin{equation}
  u_{\rm avg} = \langle | \mathbfit B - \mathbfit b | \rangle,
\end{equation}
where $\mathbfit b$ is the analytic solution, and the 
operator $\langle ||\rangle$ is the average of $||$ over the domain
and three components. As in Section \ref{sec_convergence}, $||$ 
is the componentwise absolute value. These metrics measure 
the maximum and average truncation errors respectively.\\

\noindent It should be noted that  $u_{\rm max}$ is a particularly 
strict measure of the numerical error, since a discrepancy at a single 
point can significantly affect its value. We have chosen to use this
metric as it is useful in detecting errors in the implementation
of the boundary conditions. A systematic, resolution independent
error at each of the $N^2$ points in a boundary layer introduces an 
error with scaling $\sim 1/N$ when
averaged over the whole $N^3$ points in the domain. Hence,  the
scaling of  $u_{\rm avg}$ cannot  necessarily distinguish between a method that
has $\sim 1/N$ truncation error in the volume, and a method that
systematically fails at the boundary, due to, for example, coding
errors. This is why we measure both the scaling for 
$u_{\rm max}$ and $u_{\rm avg}$ --- the metric $u_{\rm max}$ will
not decrease with $N$ if systematic errors at the boundaries
exist.\\

\begin{figure} 
  \centerline{\includegraphics[width=1.\textwidth]{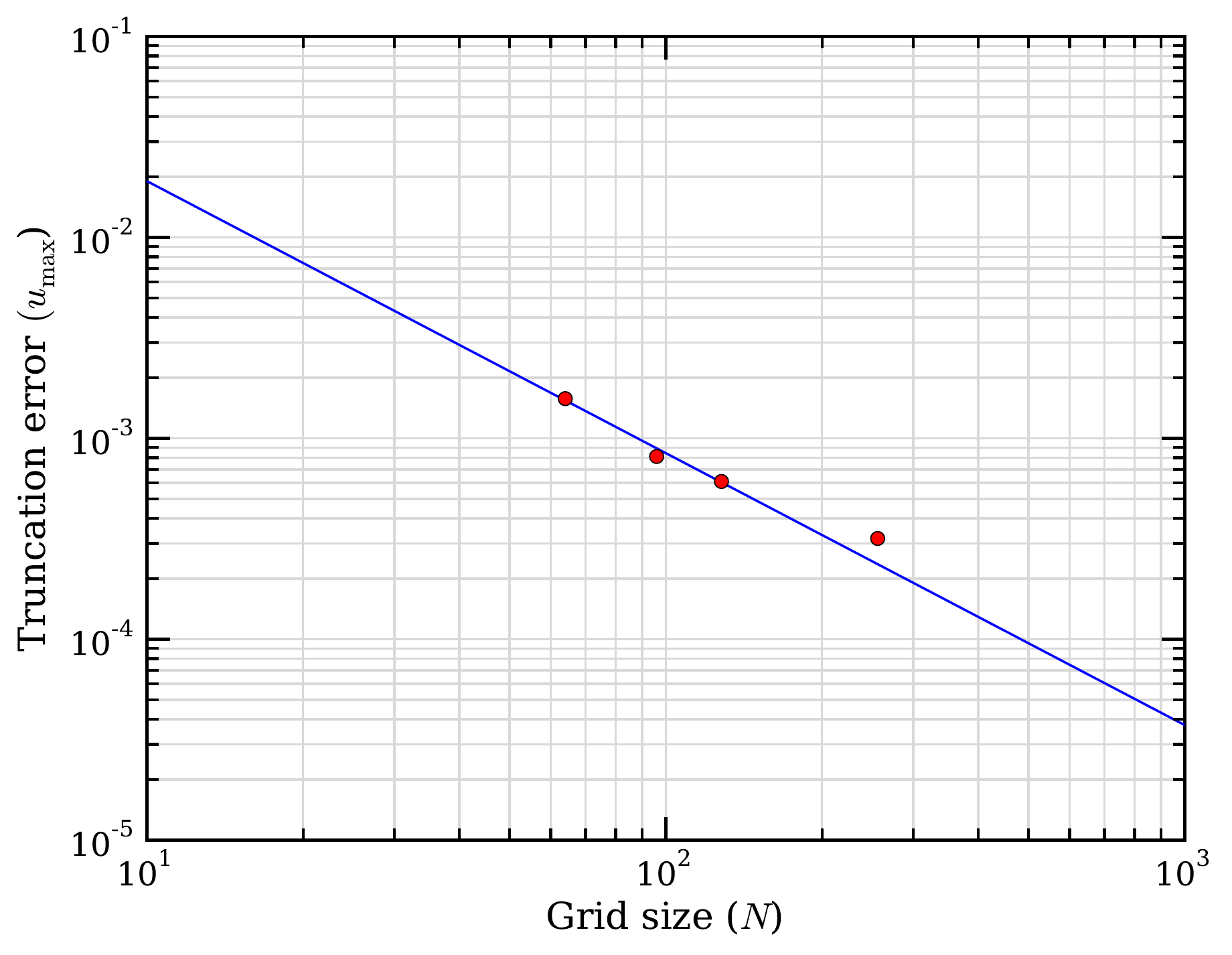}}
  \centerline{\includegraphics[width=1.\textwidth]{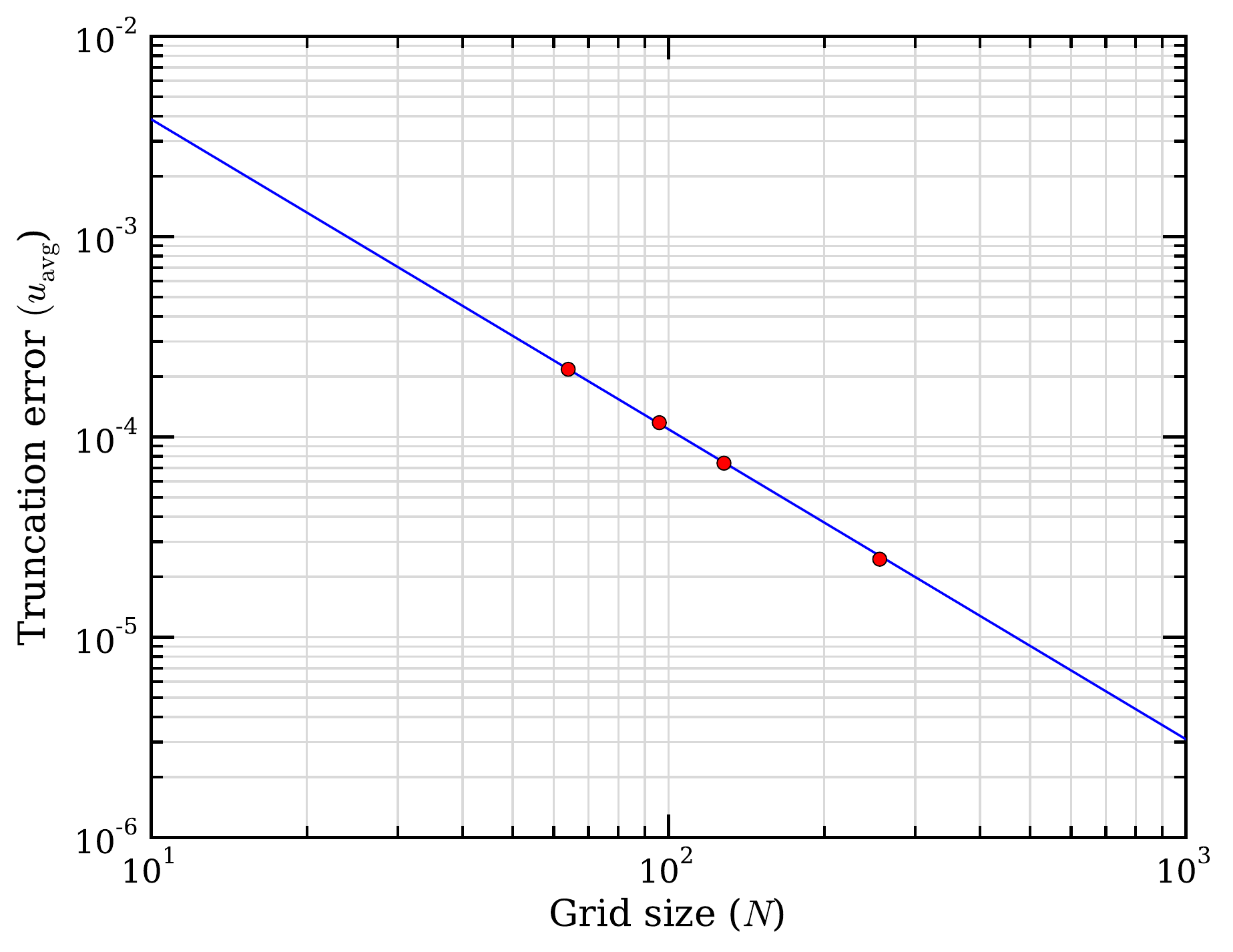}}

  \caption{Truncation error {\it versus} grid resolution as measured by the 
           $u_{\rm max}$ (top panel) and $u_{\rm avg}$ (bottom panel)
           metrics. The solid lines are power-law fits to the data
           with power-law indices $-1.35$ (top panel) and 
           $-1.55$ (bottom panel).}
  \label{fig_error}
\end{figure}

\noindent Figure \ref{fig_error} shows the truncation error {\it versus}
$N$, the number of mesh points along each dimension.
The top panel shows $u_{\rm max}$ {\it versus} $N$ (circles). 
The solid line is a power-law fit to the data with power-law index 
\iumax. The bottom panel shows $u_{\rm avg}$ {\it versus} $N$ (circles). 
The solid line shows a power-law fit with power-law index \iuavg.
Based on a visual inspection, there is good agreement between the 
fitted power-laws and the data in both panels, although the value of
$u_{\rm max}$ at $N=256$ is larger than the prediction of the fit.\\

\begin{figure} 
  \centerline{\includegraphics[width=1.\textwidth]{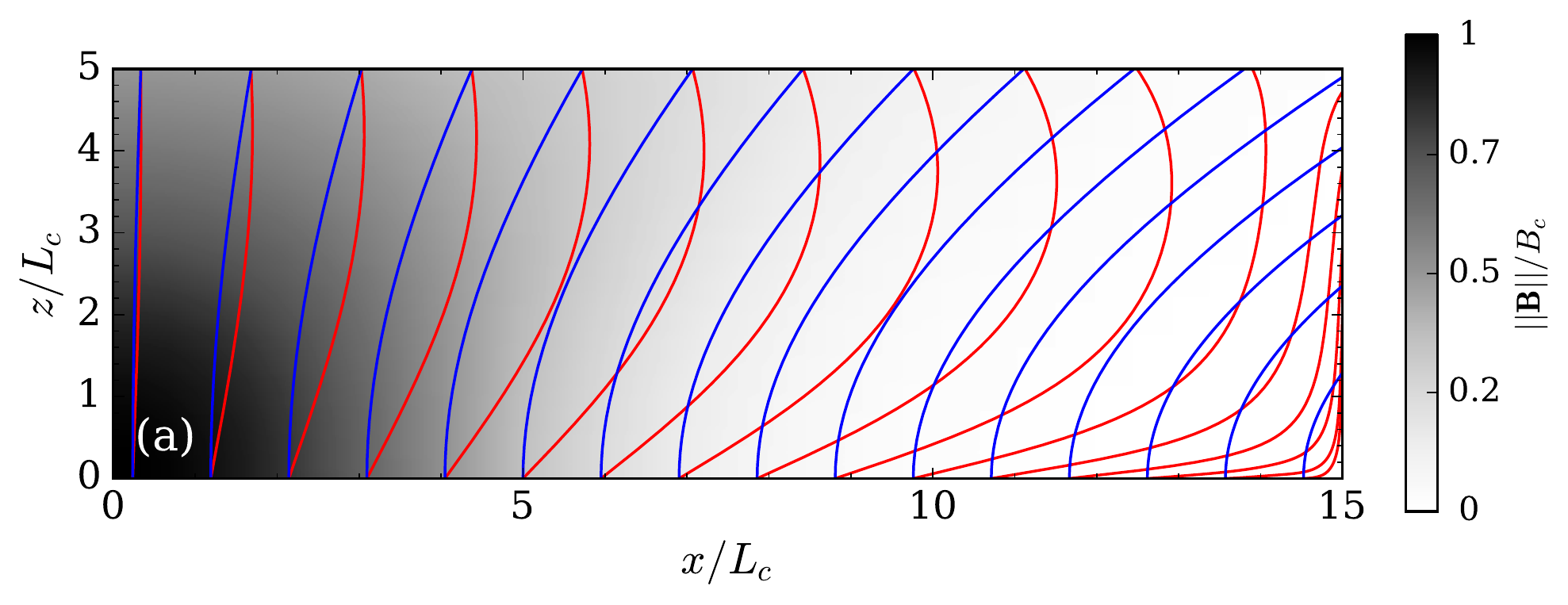}}
  \centerline{\includegraphics[width=1.\textwidth]{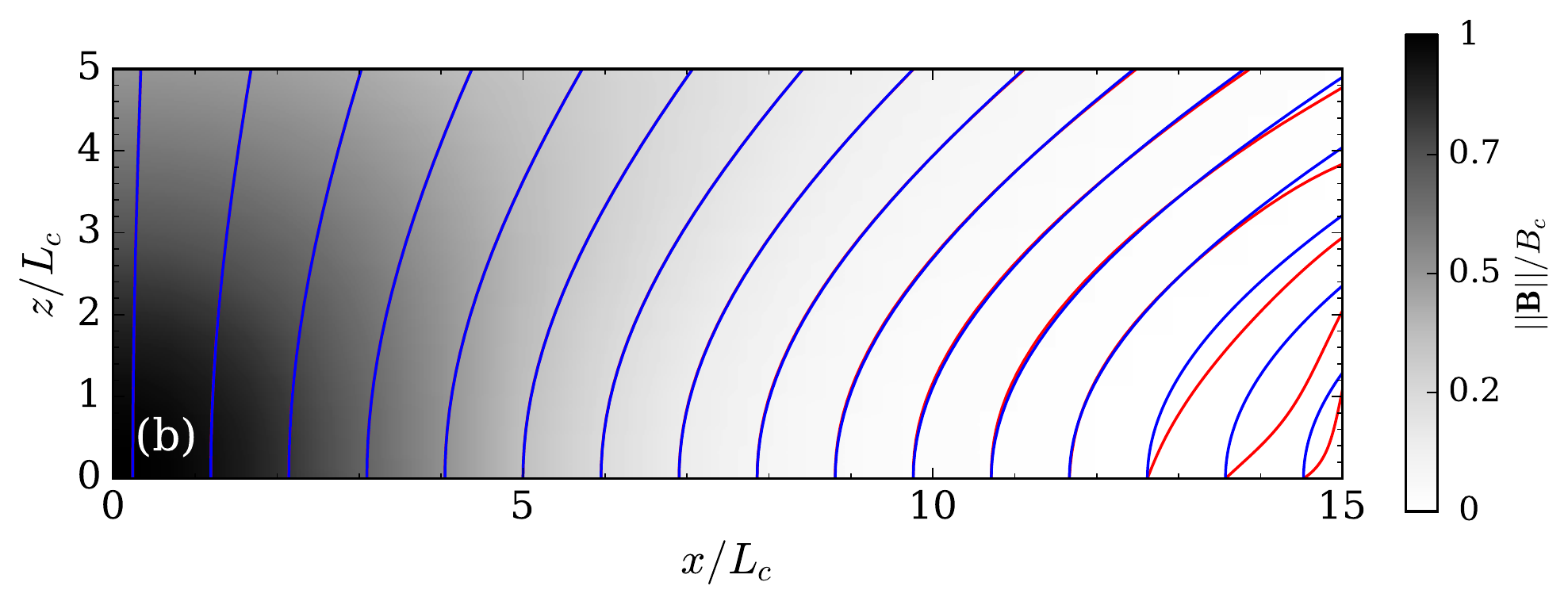}}
  
  \centerline{\includegraphics[width=1.\textwidth]{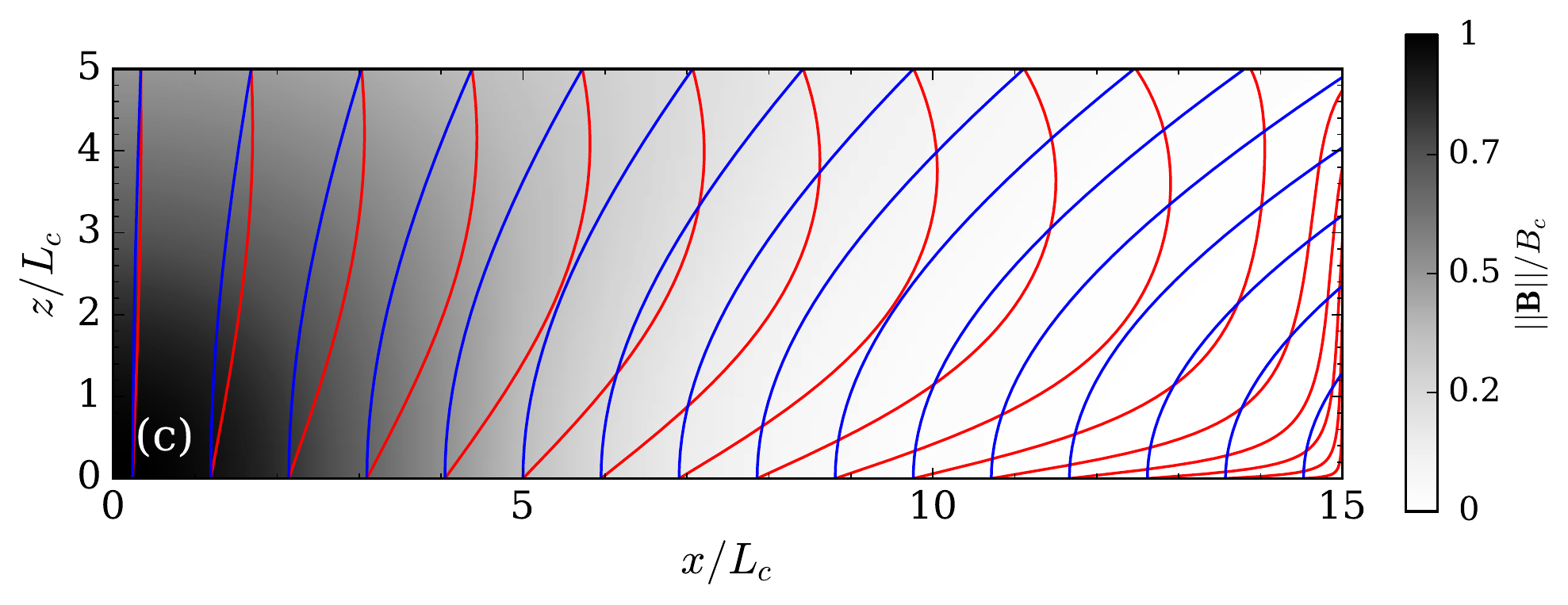}}
  \centerline{\includegraphics[width=1.\textwidth]{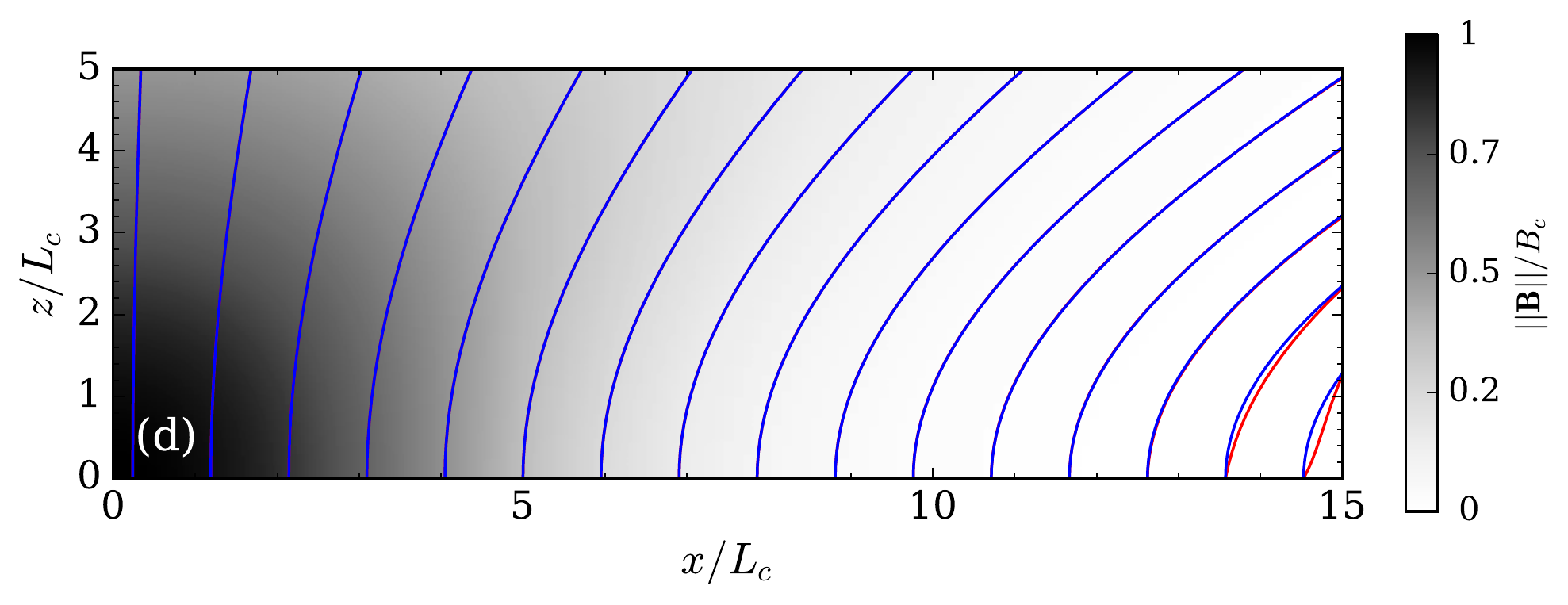}}

  \caption{
           Field lines of the exact solution (blue) superimposed on
           those of the numerical solutions (red) in the $x-z$ plane through the 
           origin. In panels (a) and (b), the red field lines belong  
           to the initial potential field and the solution calculation after
           20 iterations of the fixed-point method, respectively, for the 
           calculation with $N=64$. Panels (c) and (d) show
           the same for the $N=256$ calculation. In all panels the field lines are traced 
           from starting points spaced equally along the $x$ axis, meaning 
           the density of field lines is not indicative of field strength.
           }
  \label{fig_lines}
\end{figure}

\noindent In addition to comparing the analytic and numerical
solutions via metrics, it is important to perform a visual 
comparison between the field lines, as this gives some indication
of how the numerical error is distributed spatially. 
Figure  \ref{fig_lines} shows a comparison between the field 
lines of the exact solution (blue lines) and the numerical solution
at different resolutions and iteration numbers (red lines). Panels (a) and (c)
show the field lines of the exact solution (blue lines) {\it versus} the 
field lines of the initial potential field (red lines) for the 
$N=64$ and $N=256$ calculations, respectively. Panels (b) and (d) show the field 
lines of the exact solution {\it versus} the field lines of the numerical
solutions after 20 iterations of the method (red field lines) for
the $N=64$ and $N=256$ calculations respectively. In all panels, 
the field lines are confined to the $x-z$ plane through the origin and
only the subdomain with $x \in [0,L_x/2]$ is shown.
Since the solution is axisymmetric, the two-dimensional slice gives a 
good indication of the solution in general. The tracing is initiated from starting
points spaced equally along the $x$ axis at $z=0$. The background
image of each panel depicts $||\mathbfit B^{[k]}||$. We note that 
in the corners where the numerical and exact solutions are most 
discrepant, the magnetic field strength is smaller than that near the
center by several orders of magnitude.\\

\begin{figure} 
  \centerline{\includegraphics[width=1.\textwidth]{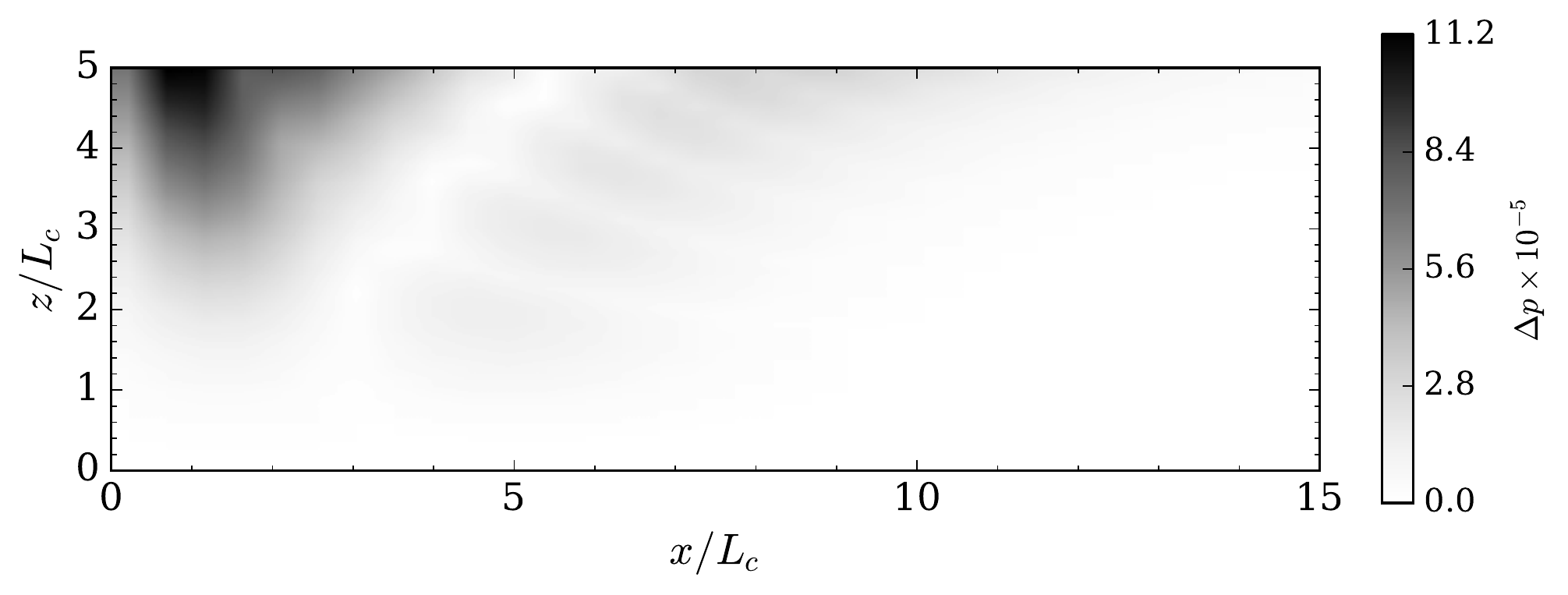}}
  \centerline{\includegraphics[width=1.\textwidth]{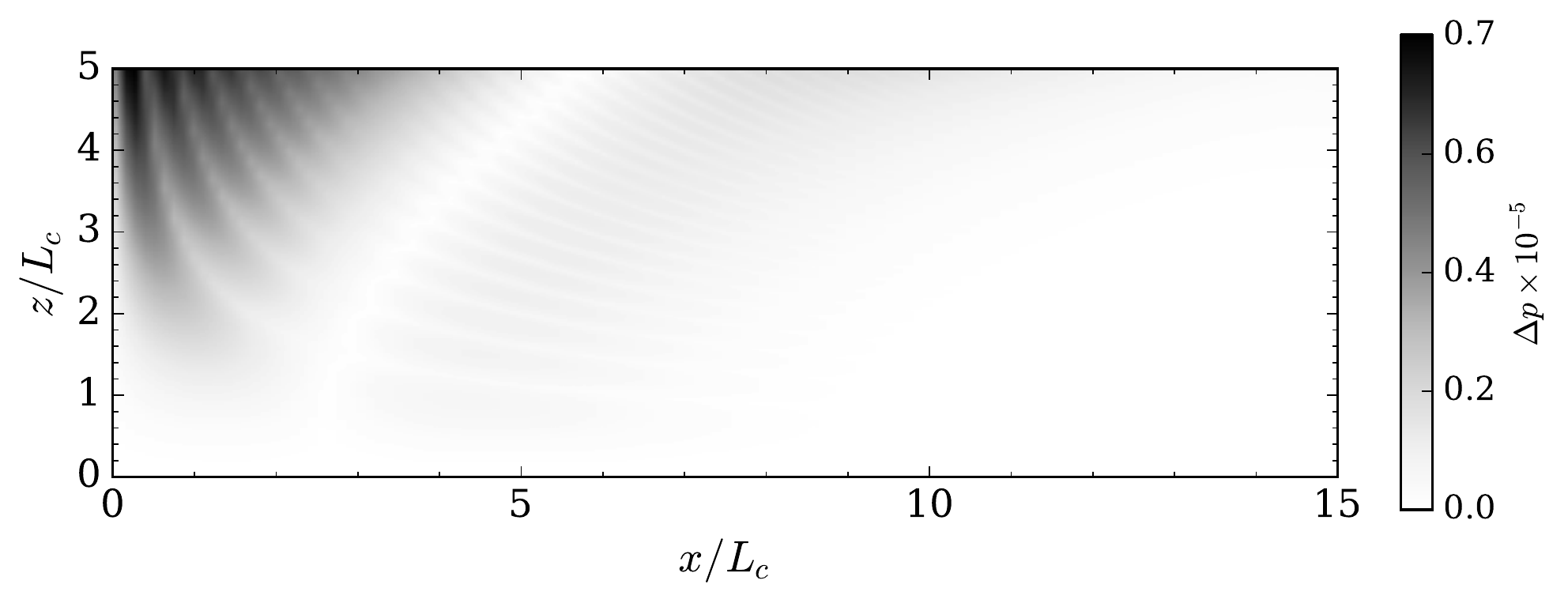}}
  
  \caption{
           The normalized absolute difference (see Equation \ref{eq_p_delta})
           between the pressure of the analytic solution
           and the numerical solution after 20 iterations of the method
           for $N=64$ (top panel) and $N=256$ (bottom panel). Note that
           the scales on the color bars are different.
           }
  \label{fig_pressure}
\end{figure}

\noindent It is also important to check the numerical error in the
thermodynamic quantities. Figure \ref{fig_pressure} shows 
maps of the quantity   
\begin{equation}
  \Delta p = \frac{|p_a - p|}{|p_a+p|},
  \label{eq_p_delta}
\end{equation}
where $p_a$ is the analytic solution, and $||$ is the absolute value.
The top panel shows $\Delta p$  for the $N=64$ case, and the bottom
panel shows $\Delta p$ for the $N=256$ case. In both panels we only show a slice 
through the $x-z$ plane as in Figure \ref{fig_lines}.


\subsection{Measurements of the Total Calculation Time}

\begin{figure} 
  \centerline{\includegraphics[width=1.\textwidth]{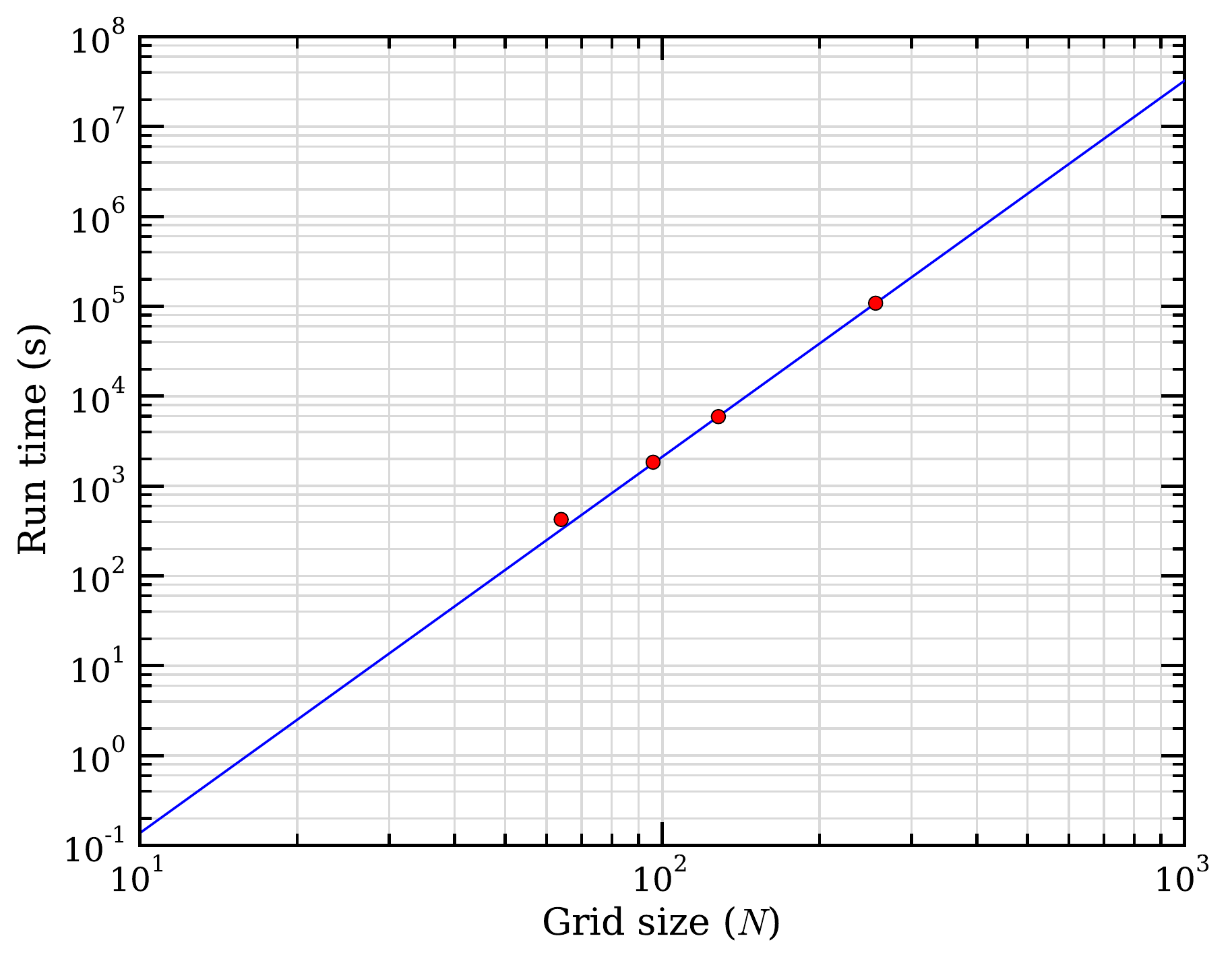}}
  
  \caption{
          Total time to complete 20 iterations of the fixed-point 
          method {\it versus} grid size $N$. The straight line
          is a power-law fit to the data with power-law index
          of $4.2$.
          }
  \label{fig_time}
\end{figure}

Finally, we compute the total time required to perform 20 iterations
of the method on an eight-core processor. Figure \ref{fig_time} shows the total run time in 
seconds {\it versus} $N$. A power-law fit to the data has a power-law
index of $4.2$ meaning the run time has approximate $\sim N^4$ 
scaling. 

%
%

\section{Discussion and Conclusion}

\noindent We present a new fixed-point iteration method for solving
the magnetohydrostatic equations in a three-dimensional Cartesian
box and its implementation in code. We apply the code to a known
analytic solution to verify that it works as expected. We perform the 
calculation at four different numerical resolutions to
determine how this affects the convergence and numerical
accuracy of the method as well as the calculation time.\\

\noindent We find that the fixed-point iteration converges in the
sense that the change in the magnetic field between iterations, as measured by
the  metric $\Delta B_{\rm max}$ (Equation (\ref{eq_con_metric})), 
decreases by approximately six orders of magnitude over 20
iterations. The decrease in   $\Delta B_{\rm max}$ appears to be exponential with a rate 
that does not depend strongly on the number of grid points. Although the 
method converged for all the cases presented, we found that 
if the resolution was made very low, by either using a small $N$
or a large domain, then the method would not reach a fixed point.
In this limit, we expect that the solution was dominated by the
numerical truncation error.\\

\noindent We also measure the numerical truncation error of the method
by comparing the known analytic solution to the numerical solution.
We do this for calculations at four different resolutions in order
to establish a scaling law for the truncation error. We find that 
for the metric $u_{\rm max}$, which is sensitive to the maximum truncation
error, the power-law index is \iumax, as determined by a fit to
the data. For the metric $u_{\rm avg}$, which is sensitive to the 
average truncation error, we find the power-law index of the fit
is \iuavg. The theoretical maximum truncation error in the field-line 
tracing solution to the hyperbolic equations is expected to 
have $\sim 1/N$ scaling (see the discussion in \citealp{2013SoPh..282..283G}), 
while the maximum truncation error for the second-order finite-difference
solution to the elliptic equations is expected to have $\sim 1/N^2$
scaling. These represent worst-case error estimates and in practice
we would expect to find a power-law index for the truncation
error somewhere in between one and two, which is what we find.
We note some departure from the fit for $u_{\rm max}$ at 
$N=256$. The value here is approximately half that at
$N=128$, so while it does not lie on the fitted line, it is still
consistent with the maximum theoretical error scaling of $\sim 1/N$.\\

\noindent We also perform a visual comparison between the field
lines of the numerical and analytic solutions. After 20 iterations,
the two sets are almost indistinguishable, except for field lines
in the lower corners of panels (b) and (d) in Figure 3. 
We found that the field lines in this region
can change significantly with even small changes in the electric
current density. The region may have the largest error because 
the analytic solution in this region departs significantly from 
the initial potential field. The magnetic field in this region 
is also very weak compared with the field at the center of the domain.
We emphasize that the discrepancy decreases with resolution (as can be seen in Figure \ref{fig_lines}), 
and would not appear to be due to the local failure of the method. \\

\noindent We also establish a scaling law for the total execution time of the
code. We find that for a grid with $N$ grid points in each dimension,
the total run time has $\sim N^4$ scaling. This scaling is consistent
with force-free codes based on the Grad-Rubin method \citep{2006SoPh..238...29W}.
As in the case of the force-free codes, the time-consuming step is
the field line tracing. We note that although the scaling is similar,
the magnetohydrostatic code is significantly slower in absolute
terms because a single iteration of the magnetohydrostatic method
involves more stages than the force-free Grad-Rubin method.\\

\noindent Our method requires that all field lines connect to the 
boundary, but no actual constraints on the connectivity
are imposed during the iteration to enforce this. In one sense this is an advantage
of the method, because it means the method can compute solutions whose topologies 
differ significantly from the magnetic field used to initiate the calculation.
However, in another sense it is a weakness, because
there is nothing to prevent closed field lines from forming during 
the calculation, at which point the calculation cannot proceed. 
We find that in our tests the appearance of closed field lines is generally
in response to the formation of strong electric currents
in weak-field regions. Future versions of the method could solve
a more general formulation of the boundary-value problem that accounts
for closed field lines. Despite this limitation, the method
is still applicable to a range of interesting problems.\\

\noindent Finally, we note that the test case used is particularly
simple. In particular, because it is untwisted, step iv) of the
method is not tested. The numerical methods are similar to 
those used to solve step \ref{p_update}), and we have tested the method 
on analytic solutions with a finite $\sigma$ but no gravity,
and we have found that the known solution is well reproduced. This gives 
us confidence that the method would work when applied to a case
with both a finite twist and gravity force. We stress that although the test case
was axisymmetric, since we work in Cartesian coordinates, our
calculation was three-dimensional.\\

\noindent The work presented here has several limitations that 
could be addressed with future work. The fixed-point method 
could be generalized to explicitly model the energetics via
an energy equation similar to the approach of \citet{1993ApJ...404..788P}.
Regarding the code itself, no significant optimization has yet been
performed on the current first version, which could be addressed 
in the future.\\

\noindent In this article we present a new iterative method for solving
the magnetohydrostatic equations in a three-dimensional Cartesian
domain and the details of an implementation of the method in code. 
We use our code to reconstruct a known analytic solution and thereby 
establish the correctness of the code and the viability of the 
method in general. This work is a step towards the generation
of realistic three-dimensional magnetohydrostatic models of 
the solar atmosphere.\\

%
%

\subsection*{Appendix A}
\label{sec_stable}

\noindent In this appendix we explain why the decomposition of 
the pressure into a magnetohydrostatic and a hydrostatic components
presented in Section \ref{sec_pd_decomp} is necessary for 
the stability of the fixed-point method. We find that without
solving for these components separately, an instability
occurs due to the failure to achieve exact hydrostatic force 
balance in weak-field regions.\\

\noindent In terms of the total pressure, $p$, and density, $\rho$, 
the update equation for the perpendicular electric current density
is
\begin{equation}
   \mathbfit J_{\perp}^{[k+1]} = \frac{
                             - \nabla p^{[k+1]} \times \mathbfit B^{[k]}
                             + \rho^{[k+1]} \mathbfit g \times \mathbfit B^{[k]}
                            }{ 
                            ||\mathbfit B^{[k]}||^2
                            }.
\label{eq_Jp_norm}
\end{equation}
Using this equation rather than Equation (\ref{iter_eq3}) results
in the formation of spurious electric currents in weak-field regions,
where the denominator becomes small but the numerator remains 
finite due to numerical error. To understand this, let $p$ and
$\rho$ be split as in Section \ref{sec_pd_decomp}. 
Equation (\ref{eq_Jp_norm}) then becomes
\begin{equation}
   \mathbfit J_{\perp}^{[k+1]} = \frac{
                             (- \nabla p^{[k+1]}_{\rm mhs} \times \mathbfit B^{[k]}
                             + \rho^{[k+1]}_{\rm mhs} \mathbfit g \times \mathbfit B^{[k]})
 +
                            (- \nabla p^{[k+1]}_{\rm hs} \times \mathbfit B^{[k]}
                             + \rho^{[k+1]}_{\rm hs} \mathbfit g \times \mathbfit B^{[k]})
                            }{ 
                            ||\mathbfit B^{[k]}||^2
                            }.
\end{equation}
In principle, the second term in the numerator is zero, however,
in practice this is not achieved numerically,  which introduces
an error, $\epsilon(\mathbfit r)$, in the numerator.
The functional form of $\epsilon(\mathbfit r)$ depends on the details of the 
numerical implementation, but, in general, its magnitude 
varies with position and decreases with resolution. In 
weak-field regions $p_{\rm mhs}$ and $\rho_{\rm mhs}$ are small, and
thus the perpendicular current density has scaling $\sim \epsilon/||\mathbfit B^{[k]}||$.
The hydrostatic component of the atmosphere is independent of the 
magnetic field, and so the error $\epsilon$ is not necessarily small
in weak-field regions. This results in the formation of strong
spurious currents in weak-field regions because the denominator
$||\mathbfit B^{[k]}||$ becomes small while the numerator remains finite.
These currents can prevent the method from converging.\\

%
%

\section*{Appendix B}

\noindent Here we provide the details of the pressure update step. \\

\noindent To update the pressure at each iteration it is necessary
to solve the equations
\begin{equation}
  \frac{dp_{\rm mhs}}{ds} + F(s)p_{\rm mhs} = G(s),
  \label{eq_ap1}
\end{equation}
and
\begin{equation}
  \frac{d \mathbfit x}{ds} = -\frac{\mathbfit B}{||\mathbfit B||},
\end{equation}
in the volume at each Cartesian mesh point $(x_i,y_j,z_k)$.
The functions $F(s)$ and $G(s)$ are defined by Equations 
(\ref{eq_hyper_g}) and (\ref{eq_hyper_f}), respectively.  
The boundary conditions are 
\begin{equation}
  \left. \mathbfit x \right |_{s=0} = (x_i,y_j,z_k),
\end{equation}
and
\begin{equation}
  \left. p_{\rm mhs} \right |_{s=s_0} = p_0,
  \label{eq_bc_apa}
\end{equation}
where $x(s_0),y(s_0),z(s_0)$ is the point where the field line 
crosses the boundary with the polarity over which boundary
conditions are prescribed.\\

\noindent This is not an initial-value problem because $\mathbfit x$ and $p$ 
are not known simultaneously at either $s$ or $s_0$. 
Application of a standard iterative method for solving 
ordinary-differential equations (like the Runge-Kutta method) is therefore impossible, because the iteration cannot be 
initialized without knowledge of both $\mathbfit x$ and $p$ at the same point.
The problem, however, can be reformulated into two 
auxiliary initial-value problems that can be solved with a straightforward 
application of standard methods.\\

\noindent First consider $p_a$, which is the solution to the equation
\begin{equation}
  \frac{dp_{a}}{ds} + F(s)p_{a} = G(s),
  \label{eq_ap2}
\end{equation}
with initial condition $p_a(0) = A$. Also consider $p_b$, which is the solution of the 
homogeneous equation
\begin{equation}
  \frac{dp_{b}}{ds} + F(s)p_{b} = 0,
  \label{eq_ap3}
\end{equation}
with $p_b(0) = B$. In both cases the equation and initial condition
for $\mathbfit x$ are the same as the original problem. Note 
that $\lambda p_b$, where $\lambda$ is a scalar, is also a solution 
to Equation (\ref{eq_ap3}). It follows that, since Equation
(\ref{eq_ap1}) is linear,
\begin{equation}
  p_c = p_a + \lambda p_b, 
\end{equation}
is also solution of Equation (\ref{eq_ap1}). The variable $\lambda$ 
can be treated as a free parameter and chosen such that  $p_c = p_0$ at the
point where the field line crosses the boundary, {\it i.e.}
\begin{equation}
  \lambda = (p_0 - p_a(s_0))/p_b(s_0).
\end{equation}
With this choice, $p_c$ is a solution to the original boundary-value
problem: it satisfies Equation  (\ref{eq_ap1}) and the boundary
conditions defined by Equation (\ref{eq_bc_apa}). It follows that
the value of $p_{\rm mhs}$ at the point $s=0$, with Cartesian 
coordinates $(x_i,y_j,z_k)$, is 
\begin{equation}
  p_{\rm mhs}(0) = p_a(0) + \lambda p_b(0) = A + B\lambda. 
\end{equation}
Hence the solution to the boundary-value problem can be found by
computing $\lambda$, which requires solving the two initial-value 
problems for $p_a$ and $p_b$. The simplest choice for the
constants $A$ and $B$ is $A=0$ and $B=1$, although other
combinations are possible. The initial-value problems for $p_a$
and $p_b$ are in the standard form and can be treated using
a straightforward application of a method like Runge-Kutta.

%
%

%
%
%

%
%
%
%
%
%
%
%

%
%
%

%
%
%
%
%
%

%
%
%
%
%

%
\begin{acks}
Support for this work is provided by the NASA Living With a Star program
through grant NNX14AD42G, and by the Solar Terrestrial program of the
National Science Foundation through grant AGS-1127327. 
\end{acks}
  
%
\begin{coi}
  The authors declare that they have no conflicts of interest.
\end{coi}

%
%

\begin{thebibliography}{63}
\ifx\bisbn     \undefined \def\bisbn  #1{ISBN #1}\fi
\ifx\binits    \undefined \def\binits#1{#1}\fi
\ifx\bauthor   \undefined \def\bauthor#1{#1}\fi
\ifx\batitle   \undefined \def\batitle#1{#1}\fi
\ifx\bjtitle   \undefined \def\bjtitle#1{\textit{#1}}\fi
\ifx\bvolume   \undefined \def\bvolume#1{\textbf{#1}}\fi
\ifx\byear     \undefined \def\byear#1{#1}\fi
\ifx\bissue    \undefined \def\bissue#1{#1}\fi
\ifx\bfpage    \undefined \def\bfpage#1{#1}\fi
\ifx\blpage    \undefined \def\blpage #1{#1}\fi
\ifx\burl      \undefined \def\burl#1{\textsf{#1}}\fi
\ifx\href      \undefined \def\href#1#2{\textsf{#2}}\fi
\ifx\betal     \undefined \def\betal{\textit{et al.}}\fi
\ifx\bctitle   \undefined \def\bctitle#1{#1}\fi
\ifx\beditor   \undefined \def\beditor#1{#1}\fi
\ifx\bbtitle   \undefined \def\bbtitle#1{\textit{#1}}\fi
\ifx\bedition  \undefined \def\bedition#1{#1}\fi
\ifx\bseriesno \undefined \def\bseriesno#1{\textbf{#1}}\fi
\ifx\blocation \undefined \def\blocation#1{#1}\fi
\ifx\bsertitle \undefined \def\bsertitle#1{\textit{#1}}\fi
\ifx\bsnm      \undefined \def\bsnm#1{#1}\fi
\ifx\bsuffix   \undefined \def\bsuffix#1{#1}\fi
\ifx\bparticle \undefined \def\bparticle#1{#1}\fi
\ifx\barticle  \undefined \def\barticle#1{}\fi
\ifx\binstitute  \undefined \def\binstitute#1{#1}\fi
\ifx\bpublisher  \undefined \def\bpublisher#1{#1}\fi
\ifx\doiurl    \undefined
  \def\doiurl#1{\href{http://dx.doi.org/#1}{\textsf{DOI}}}\fi
\ifx\arxivurl  \undefined
  \def\arxivurl#1{\href{http://arxiv.org/abs/#1}{\textsf{arXiv}}}\fi
\ifx\adsurl    \undefined
  \def\adsurl#1{\href{http://adsabs.harvard.edu/abs/#1}{\textsf{ADS}}}\fi
\ifx\botherref \undefined \def\botherref#1{}\fi
\ifx\url       \undefined \def\url#1{\textsf{#1}}\fi
\ifx\bchapter  \undefined \def\bchapter#1{}\fi
\ifx\bbook     \undefined \def\bbook#1{}\fi
\ifx\bcomment  \undefined \def\bcomment#1{#1}\fi
\ifx\oauthor   \undefined \def\oauthor#1{#1}\fi
\ifx\citeauthoryear \undefined\def \citeauthoryear#1{#1}\fi
\ifx\endbibitem\undefined \def\endbibitem{}\fi
\ifx\bconflocation  \undefined \def\bconflocation#1{#1} \fi

\bibitem[\protect\citeauthoryear{Amari, Boulbe, and
  Boulmezaoud}{2009}]{doi:10.1137/070700942}
\begin{barticle}
\bauthor{\bsnm{Amari}, \binits{T.}},
\bauthor{\bsnm{Boulbe}, \binits{C.}},
\bauthor{\bsnm{Boulmezaoud}, \binits{T.Z.}}:
\byear{2009},
\batitle{Computing beltrami fields}.
\bjtitle{SIAM J. Scien. Compt.}
\bvolume{31}(\bissue{5}),
\bfpage{3217}.
\doiurl{10.1137/070700942}.
\end{barticle}
\endbibitem

\bibitem[\protect\citeauthoryear{{Amari}, {Boulmezaoud}, and
  {Mikic}}{1999}]{1999A&A...350.1051A}
\begin{barticle}
\bauthor{\bsnm{{Amari}}, \binits{T.}},
\bauthor{\bsnm{{Boulmezaoud}}, \binits{T.Z.}},
\bauthor{\bsnm{{Mikic}}, \binits{Z.}}:
\byear{1999},
\batitle{{An iterative method for the reconstructionbreak of the solar coronal
  magnetic field. I. Method for regular solutions}}.
\bjtitle{\aap}
\bvolume{350},
\bfpage{1051}.
\adsurl{1999A\%26A...350.1051A}.
\end{barticle}
\endbibitem

\bibitem[\protect\citeauthoryear{{Amari}
  \textit{et~al.}}{2013}]{2013A&A...553A..43A}
\begin{barticle}
\bauthor{\bsnm{{Amari}}, \binits{T.}},
\bauthor{\bsnm{{Aly}}, \binits{J.-J.}},
\bauthor{\bsnm{{Canou}}, \binits{A.}},
\bauthor{\bsnm{{Mikic}}, \binits{Z.}}:
\byear{2013},
\batitle{{Reconstruction of the solar coronal magnetic field in spherical
  geometry}}.
\bjtitle{\aap}
\bvolume{553},
\bfpage{A43}.
\doiurl{10.1051/0004-6361/201220787}.
\adsurl{2013A\%26A...553A..43A}.
\end{barticle}
\endbibitem

\bibitem[\protect\citeauthoryear{{Aulanier}
  \textit{et~al.}}{1998}]{1998SoPh..183..369A}
\begin{barticle}
\bauthor{\bsnm{{Aulanier}}, \binits{G.}},
\bauthor{\bsnm{{D{\'e}moulin}}, \binits{P.}},
\bauthor{\bsnm{{Schmieder}}, \binits{B.}},
\bauthor{\bsnm{{Fang}}, \binits{C.}},
\bauthor{\bsnm{{Tang}}, \binits{Y.H.}}:
\byear{1998},
\batitle{{Magnetohydrostatic Model of a Bald-Patch Flare}}.
\bjtitle{\solphys}
\bvolume{183},
\bfpage{369}.
\doiurl{10.1023/A:1005003426798}.
\adsurl{1998SoPh..183..369A}.
\end{barticle}
\endbibitem

\bibitem[\protect\citeauthoryear{{Aulanier}
  \textit{et~al.}}{1999}]{1999A&A...342..867A}
\begin{barticle}
\bauthor{\bsnm{{Aulanier}}, \binits{G.}},
\bauthor{\bsnm{{D{\'e}moulin}}, \binits{P.}},
\bauthor{\bsnm{{Mein}}, \binits{N.}},
\bauthor{\bsnm{{van Driel-Gesztelyi}}, \binits{L.}},
\bauthor{\bsnm{{Mein}}, \binits{P.}},
\bauthor{\bsnm{{Schmieder}}, \binits{B.}}:
\byear{1999},
\batitle{{3-D magnetic configurations supporting prominences. III. Evolution of
  fine structures observed in a filament channel}}.
\bjtitle{\aap}
\bvolume{342},
\bfpage{867}.
\adsurl{1999A\%26A...342..867A}.
\end{barticle}
\endbibitem

\bibitem[\protect\citeauthoryear{{Bogdan} and
  {Low}}{1986}]{1986ApJ...306..271B}
\begin{barticle}
\bauthor{\bsnm{{Bogdan}}, \binits{T.J.}},
\bauthor{\bsnm{{Low}}, \binits{B.C.}}:
\byear{1986},
\batitle{{The three-dimensional structure of magnetostatic atmospheres. II -
  Modeling the large-scale corona}}.
\bjtitle{\apj}
\bvolume{306},
\bfpage{271}.
\doiurl{10.1086/164341}.
\adsurl{1986ApJ...306..271B}.
\end{barticle}
\endbibitem

\bibitem[\protect\citeauthoryear{Briggs, Henson, and
  McCormick}{2000}]{Briggs:2000:MT:357695}
\begin{bbook}
\bauthor{\bsnm{Briggs}, \binits{W.L.}},
\bauthor{\bsnm{Henson}, \binits{V.E.}},
\bauthor{\bsnm{McCormick}, \binits{S.F.}}:
\byear{2000},
\bbtitle{A multigrid tutorial (2nd ed.)},
\bpublisher{Society for Industrial and Applied Mathematics},
\blocation{Philadelphia, PA, USA}.
\bisbn{0-89871-462-1}.
\doiurl{10.1137/1.9780898719505}.
\end{bbook}
\endbibitem

\bibitem[\protect\citeauthoryear{{Cally}}{1990}]{1990JCoPh..89..490C}
\begin{barticle}
\bauthor{\bsnm{{Cally}}, \binits{P.S.}}:
\byear{1990},
\batitle{{An inverse coordinate multigrid method for free boundary
  magnetohydrostatics}}.
\bjtitle{J. Comput. Phys.}
\bvolume{89},
\bfpage{490}.
\doiurl{10.1016/0021-9991(90)90164-V}.
\adsurl{1990JCoPh..89..490C}.
\end{barticle}
\endbibitem

\bibitem[\protect\citeauthoryear{{Cameron}
  \textit{et~al.}}{2011}]{2011SoPh..268..293C}
\begin{barticle}
\bauthor{\bsnm{{Cameron}}, \binits{R.H.}},
\bauthor{\bsnm{{Gizon}}, \binits{L.}},
\bauthor{\bsnm{{Schunker}}, \binits{H.}},
\bauthor{\bsnm{{Pietarila}}, \binits{A.}}:
\byear{2011},
\batitle{{Constructing Semi-Empirical Sunspot Models for Helioseismology}}.
\bjtitle{\solphys}
\bvolume{268},
\bfpage{293}.
\doiurl{10.1007/s11207-010-9631-3}.
\adsurl{2011SoPh..268..293C}.
\end{barticle}
\endbibitem

\bibitem[\protect\citeauthoryear{Chandra
  \textit{et~al.}}{2001}]{Chandra:2001:PPO:355074}
\begin{bbook}
\bauthor{\bsnm{Chandra}, \binits{R.}},
\bauthor{\bsnm{Dagum}, \binits{L.}},
\bauthor{\bsnm{Kohr}, \binits{D.}},
\bauthor{\bsnm{Maydan}, \binits{D.}},
\bauthor{\bsnm{McDonald}, \binits{J.}},
\bauthor{\bsnm{Menon}, \binits{R.}}:
\byear{2001},
\bbtitle{Parallel programming in openmp},
\bpublisher{Morgan Kaufmann Publishers Inc.},
\blocation{San Francisco, CA, USA}.
\bisbn{1-55860-671-8, 9781558606715}.
\end{bbook}
\endbibitem

\bibitem[\protect\citeauthoryear{Chodura and
  Schl{\"u}ter}{1981}]{CHODURA198168}
\begin{barticle}
\bauthor{\bsnm{Chodura}, \binits{R.}},
\bauthor{\bsnm{Schl{\"u}ter}, \binits{A.}}:
\byear{1981},
\batitle{A 3d code for mhd equilibrium and stability}.
\bjtitle{Journal of Computational Physics}
\bvolume{41}(\bissue{1}),
\bfpage{68 }.
\doiurl{10.1016/0021-9991(81)90080-2}.
\end{barticle}
\endbibitem

\bibitem[\protect\citeauthoryear{{Deinzer}
  \textit{et~al.}}{1984}]{1984A&A...139..426D}
\begin{barticle}
\bauthor{\bsnm{{Deinzer}}, \binits{W.}},
\bauthor{\bsnm{{Hensler}}, \binits{G.}},
\bauthor{\bsnm{{Schuessler}}, \binits{M.}},
\bauthor{\bsnm{{Weisshaar}}, \binits{E.}}:
\byear{1984},
\batitle{{Model calculations of magnetic flux tubes. I - Equations and method.
  II - Stationary results for solar magnetic elements}}.
\bjtitle{\aap}
\bvolume{139},
\bfpage{426}.
\adsurl{1984A\%26A...139..426D}.
\end{barticle}
\endbibitem

\bibitem[\protect\citeauthoryear{{DeRosa}
  \textit{et~al.}}{2009}]{2009ApJ...696.1780D}
\begin{barticle}
\bauthor{\bsnm{{DeRosa}}, \binits{M.L.}},
\bauthor{\bsnm{{Schrijver}}, \binits{C.J.}},
\bauthor{\bsnm{{Barnes}}, \binits{G.}},
\bauthor{\bsnm{{Leka}}, \binits{K.D.}},
\bauthor{\bsnm{{Lites}}, \binits{B.W.}},
\bauthor{\bsnm{{Aschwanden}}, \binits{M.J.}},
\bauthor{\bsnm{{Amari}}, \binits{T.}},
\bauthor{\bsnm{{Canou}}, \binits{A.}},
\bauthor{\bsnm{{McTiernan}}, \binits{J.M.}},
\bauthor{\bsnm{{R{\'e}gnier}}, \binits{S.}},
\bauthor{\bsnm{{Thalmann}}, \binits{J.K.}},
\bauthor{\bsnm{{Valori}}, \binits{G.}},
\bauthor{\bsnm{{Wheatland}}, \binits{M.S.}},
\bauthor{\bsnm{{Wiegelmann}}, \binits{T.}},
\bauthor{\bsnm{{Cheung}}, \binits{M.C.M.}},
\bauthor{\bsnm{{Conlon}}, \binits{P.A.}},
\bauthor{\bsnm{{Fuhrmann}}, \binits{M.}},
\bauthor{\bsnm{{Inhester}}, \binits{B.}},
\bauthor{\bsnm{{Tadesse}}, \binits{T.}}:
\byear{2009},
\batitle{{A Critical Assessment of Nonlinear Force-Free Field Modeling of the
  Solar Corona for Active Region 10953}}.
\bjtitle{\apj}
\bvolume{696},
\bfpage{1780}.
\doiurl{10.1088/0004-637X/696/2/1780}.
\adsurl{2009ApJ...696.1780D}.
\end{barticle}
\endbibitem

\bibitem[\protect\citeauthoryear{{DeRosa}
  \textit{et~al.}}{2015}]{2015ApJ...811..107D}
\begin{barticle}
\bauthor{\bsnm{{DeRosa}}, \binits{M.L.}},
\bauthor{\bsnm{{Wheatland}}, \binits{M.S.}},
\bauthor{\bsnm{{Leka}}, \binits{K.D.}},
\bauthor{\bsnm{{Barnes}}, \binits{G.}},
\bauthor{\bsnm{{Amari}}, \binits{T.}},
\bauthor{\bsnm{{Canou}}, \binits{A.}},
\bauthor{\bsnm{{Gilchrist}}, \binits{S.A.}},
\bauthor{\bsnm{{Thalmann}}, \binits{J.K.}},
\bauthor{\bsnm{{Valori}}, \binits{G.}},
\bauthor{\bsnm{{Wiegelmann}}, \binits{T.}},
\bauthor{\bsnm{{Schrijver}}, \binits{C.J.}},
\bauthor{\bsnm{{Malanushenko}}, \binits{A.}},
\bauthor{\bsnm{{Sun}}, \binits{X.}},
\bauthor{\bsnm{{R{\'e}gnier}}, \binits{S.}}:
\byear{2015},
\batitle{{The Influence of Spatial resolution on Nonlinear Force-free
  Modeling}}.
\bjtitle{\apj}
\bvolume{811},
\bfpage{107}.
\doiurl{10.1088/0004-637X/811/2/107}.
\adsurl{2015ApJ...811..107D}.
\end{barticle}
\endbibitem

\bibitem[\protect\citeauthoryear{{Fiedler} and
  {Cally}}{1990}]{1990SoPh..126...69F}
\begin{barticle}
\bauthor{\bsnm{{Fiedler}}, \binits{R.A.S.}},
\bauthor{\bsnm{{Cally}}, \binits{P.S.}}:
\byear{1990},
\batitle{{Force and energy balance in the transition region network}}.
\bjtitle{\solphys}
\bvolume{126},
\bfpage{69}.
\doiurl{10.1007/BF00158299}.
\adsurl{1990SoPh..126...69F}.
\end{barticle}
\endbibitem

\bibitem[\protect\citeauthoryear{{Gary}}{2001}]{2001SoPh..203...71G}
\begin{barticle}
\bauthor{\bsnm{{Gary}}, \binits{G.A.}}:
\byear{2001},
\batitle{{Plasma Beta above a Solar Active Region: Rethinking the Paradigm}}.
\bjtitle{\solphys}
\bvolume{203},
\bfpage{71}.
\doiurl{10.1023/A:1012722021820}.
\adsurl{2001SoPh..203...71G}.
\end{barticle}
\endbibitem

\bibitem[\protect\citeauthoryear{{Gilchrist} and
  {Wheatland}}{2013}]{2013SoPh..282..283G}
\begin{barticle}
\bauthor{\bsnm{{Gilchrist}}, \binits{S.A.}},
\bauthor{\bsnm{{Wheatland}}, \binits{M.S.}}:
\byear{2013},
\batitle{{A Magnetostatic Grad-Rubin Code for Coronal Magnetic Field
  Extrapolations}}.
\bjtitle{\solphys}
\bvolume{282},
\bfpage{283}.
\doiurl{10.1007/s11207-012-0144-0}.
\adsurl{2013SoPh..282..283G}.
\end{barticle}
\endbibitem

\bibitem[\protect\citeauthoryear{{Gilchrist} and
  {Wheatland}}{2014}]{2014SoPh..289.1153G}
\begin{barticle}
\bauthor{\bsnm{{Gilchrist}}, \binits{S.A.}},
\bauthor{\bsnm{{Wheatland}}, \binits{M.S.}}:
\byear{2014},
\batitle{{Nonlinear Force-Free Modeling of the Corona in Spherical
  Coordinates}}.
\bjtitle{\solphys}
\bvolume{289},
\bfpage{1153}.
\doiurl{10.1007/s11207-013-0406-5}.
\adsurl{2014SoPh..289.1153G}.
\end{barticle}
\endbibitem

\bibitem[\protect\citeauthoryear{{Gilchrist}, {Wheatland}, and
  {Leka}}{2012}]{2012SoPh..276..133G}
\begin{barticle}
\bauthor{\bsnm{{Gilchrist}}, \binits{S.A.}},
\bauthor{\bsnm{{Wheatland}}, \binits{M.S.}},
\bauthor{\bsnm{{Leka}}, \binits{K.D.}}:
\byear{2012},
\batitle{{The Free Energy of NOAA Solar Active Region AR 11029}}.
\bjtitle{\solphys}
\bvolume{276},
\bfpage{133}.
\doiurl{10.1007/s11207-011-9878-3}.
\adsurl{2012SoPh..276..133G}.
\end{barticle}
\endbibitem

\bibitem[\protect\citeauthoryear{Grad and Rubin}{1958}]{24756}
\begin{bchapter}
\bauthor{\bsnm{Grad}, \binits{H.}},
\bauthor{\bsnm{Rubin}, \binits{H.}}:
\byear{1958},
\bctitle{Hydromagnetic equilibria and force-free fields}.
In: \beditor{\bsnm{Martens}, \binits{J.H.}},
\beditor{\bsnm{Ourom}, \binits{L.}},
\beditor{\bsnm{Barss}, \binits{W.M.}},
\beditor{\bsnm{Bassett}, \binits{L.G.}},
\beditor{\bsnm{Smith}, \binits{K.R.E.}},
\beditor{\bsnm{Gerrard}, \binits{M.}},
\beditor{\bsnm{Hudswell}, \binits{F.}},
\beditor{\bsnm{Guttman}, \binits{B.}},
\beditor{\bsnm{Pomeroy}, \binits{J.H.}},
\beditor{\bsnm{Woollen}, \binits{W.B.}},
\beditor{\bsnm{Singwi}, \binits{K.S.}},
\beditor{\bsnm{Carr}, \binits{T.E.F.}},
\beditor{\bsnm{Kolb}, \binits{A.C.}},
\beditor{\bsnm{Matterson}, \binits{A.H.S.}},
\beditor{\bsnm{Welgos}, \binits{S.P.}},
\beditor{\bsnm{Rojanski}, \binits{I.D.}},
\beditor{\bsnm{Finkelstein}, \binits{D.}} (eds.)
\bbtitle{Peaceful Uses of Atomic Energy: Theoretical and Experimental Aspects
  of Controlled Nuclear Fusion}
\bseriesno{31},
\bpublisher{United Nations},
\blocation{Geneva},
\bfpage{190}.
\end{bchapter}
\endbibitem

\bibitem[\protect\citeauthoryear{Greene and Johnson}{1961}]{greenpaper}
\begin{barticle}
\bauthor{\bsnm{Greene}, \binits{J.M.}},
\bauthor{\bsnm{Johnson}, \binits{J.L.}}:
\byear{1961},
\batitle{Determination of hydromagnetic equilibria}.
\bjtitle{Physics of Fluids}
\bvolume{4}(\bissue{7}),
\bfpage{875}.
\doiurl{10.1063/1.1706420}.
\end{barticle}
\endbibitem

\bibitem[\protect\citeauthoryear{Hairer, N{\o}rsett, and
  Wanner}{1993}]{Hairer:1993:SOD:153158}
\begin{bbook}
\bauthor{\bsnm{Hairer}, \binits{E.}},
\bauthor{\bsnm{N{\o}rsett}, \binits{S.P.}},
\bauthor{\bsnm{Wanner}, \binits{G.}}:
\byear{1993},
\bbtitle{Solving ordinary differential equations i (2nd revised. ed.): Nonstiff
  problems},
\bpublisher{Springer},
\blocation{New York, NY, USA},
\bfpage{195}.
\bisbn{0-387-56670-8}.
\doiurl{10.1007/978-3-540-78862-1}.
\end{bbook}
\endbibitem

\bibitem[\protect\citeauthoryear{{Hennig} and
  {Cally}}{2001}]{2001SoPh..201..289H}
\begin{barticle}
\bauthor{\bsnm{{Hennig}}, \binits{B.S.}},
\bauthor{\bsnm{{Cally}}, \binits{P.S.}}:
\byear{2001},
\batitle{{Numerical Solutions of Three-Dimensional Pressure-Bounded
  Magnetohydrostatic Flux Tubes}}.
\bjtitle{\solphys}
\bvolume{201},
\bfpage{289}.
\doiurl{10.1023/A:1017574714036}.
\adsurl{2001SoPh..201..289H}.
\end{barticle}
\endbibitem

\bibitem[\protect\citeauthoryear{{Khomenko} and
  {Collados}}{2006}]{2006ApJ...653..739K}
\begin{barticle}
\bauthor{\bsnm{{Khomenko}}, \binits{E.}},
\bauthor{\bsnm{{Collados}}, \binits{M.}}:
\byear{2006},
\batitle{{Numerical Modeling of Magnetohydrodynamic Wave Propagation and
  Refraction in Sunspots}}.
\bjtitle{\apj}
\bvolume{653},
\bfpage{739}.
\doiurl{10.1086/507760}.
\adsurl{2006ApJ...653..739K}.
\end{barticle}
\endbibitem

\bibitem[\protect\citeauthoryear{{Khomenko}, {Collados}, and
  {Felipe}}{2008}]{2008SoPh..251..589K}
\begin{barticle}
\bauthor{\bsnm{{Khomenko}}, \binits{E.}},
\bauthor{\bsnm{{Collados}}, \binits{M.}},
\bauthor{\bsnm{{Felipe}}, \binits{T.}}:
\byear{2008},
\batitle{{Nonlinear Numerical Simulations of Magneto-Acoustic Wave Propagation
  in Small-Scale Flux Tubes}}.
\bjtitle{\solphys}
\bvolume{251},
\bfpage{589}.
\doiurl{10.1007/s11207-008-9133-8}.
\adsurl{2008SoPh..251..589K}.
\end{barticle}
\endbibitem

\bibitem[\protect\citeauthoryear{{Kippenhahn} and
  {Schl{\"u}ter}}{1957}]{1957ZA.....43...36K}
\begin{barticle}
\bauthor{\bsnm{{Kippenhahn}}, \binits{R.}},
\bauthor{\bsnm{{Schl{\"u}ter}}, \binits{A.}}:
\byear{1957},
\batitle{{Eine Theorie der solaren Filamente. Mit 7 Textabbildungen}}.
\bjtitle{\zap}
\bvolume{43},
\bfpage{36}.
\adsurl{1957ZA.....43...36K}.
\end{barticle}
\endbibitem

\bibitem[\protect\citeauthoryear{{Low}}{1980}]{1980SoPh...67...57L}
\begin{barticle}
\bauthor{\bsnm{{Low}}, \binits{B.C.}}:
\byear{1980},
\batitle{{Exact static equilibrium of vertically oriented magnetic flux tubes.
  I - The Schlueter-Temesvary sunspot}}.
\bjtitle{\solphys}
\bvolume{67},
\bfpage{57}.
\doiurl{10.1007/BF00146682}.
\adsurl{1980SoPh...67...57L}.
\end{barticle}
\endbibitem

\bibitem[\protect\citeauthoryear{{Low}}{1984}]{1984ApJ...277..415L}
\begin{barticle}
\bauthor{\bsnm{{Low}}, \binits{B.C.}}:
\byear{1984},
\batitle{{Three-dimensional magnetostatic atmospheres - Magnetic field with
  vertically oriented tension force}}.
\bjtitle{\apj}
\bvolume{277},
\bfpage{415}.
\doiurl{10.1086/161708}.
\adsurl{1984ApJ...277..415L}.
\end{barticle}
\endbibitem

\bibitem[\protect\citeauthoryear{{Low}}{1985}]{1985ApJ...293...31L}
\begin{barticle}
\bauthor{\bsnm{{Low}}, \binits{B.C.}}:
\byear{1985},
\batitle{{Three-dimensional structures of magnetostatic atmospheres. I -
  Theory}}.
\bjtitle{\apj}
\bvolume{293},
\bfpage{31}.
\doiurl{10.1086/163211}.
\adsurl{1985ApJ...293...31L}.
\end{barticle}
\endbibitem

\bibitem[\protect\citeauthoryear{{Low}}{1992}]{1992ApJ...399..300L}
\begin{barticle}
\bauthor{\bsnm{{Low}}, \binits{B.C.}}:
\byear{1992},
\batitle{{Three-dimensional structures of magnetostatic atmospheres. IV -
  Magnetic structures over a solar active region}}.
\bjtitle{\apj}
\bvolume{399},
\bfpage{300}.
\doiurl{10.1086/171925}.
\adsurl{1992ApJ...399..300L}.
\end{barticle}
\endbibitem

\bibitem[\protect\citeauthoryear{Metcalf, Reid, and
  Cohen}{2011}]{Metcalf:2011:MFE:2090092}
\begin{bbook}
\bauthor{\bsnm{Metcalf}, \binits{M.}},
\bauthor{\bsnm{Reid}, \binits{J.}},
\bauthor{\bsnm{Cohen}, \binits{M.}}:
\byear{2011},
\bbtitle{Modern fortran explained},
\bedition{4th} edn.
\bpublisher{Oxford University Press, Inc.},
\blocation{New York, NY, USA}.
\bisbn{0199601410, 9780199601417}.
\end{bbook}
\endbibitem

\bibitem[\protect\citeauthoryear{{Metcalf}
  \textit{et~al.}}{1995}]{1995ApJ...439..474M}
\begin{barticle}
\bauthor{\bsnm{{Metcalf}}, \binits{T.R.}},
\bauthor{\bsnm{{Jiao}}, \binits{L.}},
\bauthor{\bsnm{{McClymont}}, \binits{A.N.}},
\bauthor{\bsnm{{Canfield}}, \binits{R.C.}},
\bauthor{\bsnm{{Uitenbroek}}, \binits{H.}}:
\byear{1995},
\batitle{{Is the solar chromospheric magnetic field force-free?}}
\bjtitle{\apj}
\bvolume{439},
\bfpage{474}.
\doiurl{10.1086/175188}.
\adsurl{1995ApJ...439..474M}.
\end{barticle}
\endbibitem

\bibitem[\protect\citeauthoryear{{Moradi}, {Hanasoge}, and
  {Cally}}{2009}]{2009ApJ...690L..72M}
\begin{barticle}
\bauthor{\bsnm{{Moradi}}, \binits{H.}},
\bauthor{\bsnm{{Hanasoge}}, \binits{S.M.}},
\bauthor{\bsnm{{Cally}}, \binits{P.S.}}:
\byear{2009},
\batitle{{Numerical Models of Travel-Time Inhomogeneities in Sunspots}}.
\bjtitle{\apjl}
\bvolume{690},
\bfpage{L72}.
\doiurl{10.1088/0004-637X/690/1/L72}.
\adsurl{2009ApJ...690L..72M}.
\end{barticle}
\endbibitem

\bibitem[\protect\citeauthoryear{{Moradi}
  \textit{et~al.}}{2010}]{2010SoPh..267....1M}
\begin{barticle}
\bauthor{\bsnm{{Moradi}}, \binits{H.}},
\bauthor{\bsnm{{Baldner}}, \binits{C.}},
\bauthor{\bsnm{{Birch}}, \binits{A.C.}},
\bauthor{\bsnm{{Braun}}, \binits{D.C.}},
\bauthor{\bsnm{{Cameron}}, \binits{R.H.}},
\bauthor{\bsnm{{Duvall}}, \binits{T.L.}},
\bauthor{\bsnm{{Gizon}}, \binits{L.}},
\bauthor{\bsnm{{Haber}}, \binits{D.}},
\bauthor{\bsnm{{Hanasoge}}, \binits{S.M.}},
\bauthor{\bsnm{{Hindman}}, \binits{B.W.}},
\bauthor{\bsnm{{Jackiewicz}}, \binits{J.}},
\bauthor{\bsnm{{Khomenko}}, \binits{E.}},
\bauthor{\bsnm{{Komm}}, \binits{R.}},
\bauthor{\bsnm{{Rajaguru}}, \binits{P.}},
\bauthor{\bsnm{{Rempel}}, \binits{M.}},
\bauthor{\bsnm{{Roth}}, \binits{M.}},
\bauthor{\bsnm{{Schlichenmaier}}, \binits{R.}},
\bauthor{\bsnm{{Schunker}}, \binits{H.}},
\bauthor{\bsnm{{Spruit}}, \binits{H.C.}},
\bauthor{\bsnm{{Strassmeier}}, \binits{K.G.}},
\bauthor{\bsnm{{Thompson}}, \binits{M.J.}},
\bauthor{\bsnm{{Zharkov}}, \binits{S.}}:
\byear{2010},
\batitle{{Modeling the Subsurface Structure of Sunspots}}.
\bjtitle{\solphys}
\bvolume{267},
\bfpage{1}.
\doiurl{10.1007/s11207-010-9630-4}.
\adsurl{2010SoPh..267....1M}.
\end{barticle}
\endbibitem

\bibitem[\protect\citeauthoryear{{Neukirch}}{1997}]{1997A&A...325..847N}
\begin{barticle}
\bauthor{\bsnm{{Neukirch}}, \binits{T.}}:
\byear{1997},
\batitle{{Nonlinear self-consistent three-dimensional arcade-like solutions of
  the magnetohydrostatic equations.}}
\bjtitle{\aap}
\bvolume{325},
\bfpage{847}.
\adsurl{1997A\%26A...325..847N}.
\end{barticle}
\endbibitem

\bibitem[\protect\citeauthoryear{{Petrie}}{2000}]{2000PhDT.........2P}
\begin{botherref}
\oauthor{\bsnm{{Petrie}}, \binits{G.J.D.}}:
2000,
{Three-dimensional Equilibrium Solutions to the Magnetohydrodynamic Equations
  and their Application to Solar Coronal Structures}.
PhD thesis,
Univ. of St.~Andrews.
\adsurl{2000PhDT.........2P}.
\end{botherref}
\endbibitem

\bibitem[\protect\citeauthoryear{{Petrie} and
  {Neukirch}}{2000}]{2000A&A...356..735P}
\begin{barticle}
\bauthor{\bsnm{{Petrie}}, \binits{G.J.D.}},
\bauthor{\bsnm{{Neukirch}}, \binits{T.}}:
\byear{2000},
\batitle{{The Green's function method for a special class of linear
  three-dimensional magnetohydrostatic equilibria}}.
\bjtitle{\aap}
\bvolume{356},
\bfpage{735}.
\adsurl{2000A\%26A...356..735P}.
\end{barticle}
\endbibitem

\bibitem[\protect\citeauthoryear{{Pizzo}}{1986}]{1986ApJ...302..785P}
\begin{barticle}
\bauthor{\bsnm{{Pizzo}}, \binits{V.J.}}:
\byear{1986},
\batitle{{Numerical solution of the magnetostatic equations for thick flux
  tubes, with application to sunspots, pores, and related structures}}.
\bjtitle{\apj}
\bvolume{302},
\bfpage{785}.
\doiurl{10.1086/164041}.
\adsurl{1986ApJ...302..785P}.
\end{barticle}
\endbibitem

\bibitem[\protect\citeauthoryear{{Pizzo}}{1990}]{1990ApJ...365..764P}
\begin{barticle}
\bauthor{\bsnm{{Pizzo}}, \binits{V.J.}}:
\byear{1990},
\batitle{{Numerical modeling of solar magnetostatic structures bounded by
  current sheets}}.
\bjtitle{\apj}
\bvolume{365},
\bfpage{764}.
\doiurl{10.1086/169530}.
\adsurl{1990ApJ...365..764P}.
\end{barticle}
\endbibitem

\bibitem[\protect\citeauthoryear{{Pizzo}, {MacGregor}, and
  {Kunasz}}{1993}]{1993ApJ...404..788P}
\begin{barticle}
\bauthor{\bsnm{{Pizzo}}, \binits{V.J.}},
\bauthor{\bsnm{{MacGregor}}, \binits{K.B.}},
\bauthor{\bsnm{{Kunasz}}, \binits{P.B.}}:
\byear{1993},
\batitle{{A numerical simulation of two-dimensional radiative equilibrium in
  magnetostatic flux tubes. I - The model}}.
\bjtitle{\apj}
\bvolume{404},
\bfpage{788}.
\doiurl{10.1086/172333}.
\adsurl{1993ApJ...404..788P}.
\end{barticle}
\endbibitem

\bibitem[\protect\citeauthoryear{{Pneuman} and
  {Kopp}}{1971}]{1971SoPh...18..258P}
\begin{barticle}
\bauthor{\bsnm{{Pneuman}}, \binits{G.W.}},
\bauthor{\bsnm{{Kopp}}, \binits{R.A.}}:
\byear{1971},
\batitle{{Gas-Magnetic Field Interactions in the Solar Corona}}.
\bjtitle{\solphys}
\bvolume{18},
\bfpage{258}.
\doiurl{10.1007/BF00145940}.
\adsurl{1971SoPh...18..258P}.
\end{barticle}
\endbibitem

\bibitem[\protect\citeauthoryear{Press
  \textit{et~al.}}{2007}]{Press:2007:NRE:1403886}
\begin{bbook}
\bauthor{\bsnm{Press}, \binits{W.H.}},
\bauthor{\bsnm{Teukolsky}, \binits{S.A.}},
\bauthor{\bsnm{Vetterling}, \binits{W.T.}},
\bauthor{\bsnm{Flannery}, \binits{B.P.}}:
\byear{2007},
\bbtitle{Numerical recipes 3rd edition: The art of scientific computing},
\bedition{3}rd edn.
\bpublisher{Cambridge University Press},
\blocation{New York, NY, USA},
\bfpage{150}.
\bisbn{0521880688, 9780521880688}.
\end{bbook}
\endbibitem

\bibitem[\protect\citeauthoryear{{Priest}}{2014}]{9559}
\begin{bbook}
\bauthor{\bsnm{{Priest}}, \binits{E.}}:
\byear{2014},
\bbtitle{Magnetohydrodynamics of the sun},
\bpublisher{Cambridge University Press},
\blocation{Cambridge, UK},
\bfpage{133}.
\bisbn{9781139020732}.
\doiurl{10.1017/CBO9781139020732}.
\adsurl{2014masu.book.....P}.
\end{bbook}
\endbibitem

\bibitem[\protect\citeauthoryear{{R{\'e}gnier}, {Amari}, and
  {Kersal{\'e}}}{2002}]{2002A&A...392.1119R}
\begin{barticle}
\bauthor{\bsnm{{R{\'e}gnier}}, \binits{S.}},
\bauthor{\bsnm{{Amari}}, \binits{T.}},
\bauthor{\bsnm{{Kersal{\'e}}}, \binits{E.}}:
\byear{2002},
\batitle{{3D Coronal magnetic field from vector magnetograms:
  non-constant-alpha force-free configuration of the active region NOAA 8151}}.
\bjtitle{\aap}
\bvolume{392},
\bfpage{1119}.
\doiurl{10.1051/0004-6361:20020993}.
\adsurl{2002A\%26A...392.1119R}.
\end{barticle}
\endbibitem

\bibitem[\protect\citeauthoryear{{Ruan}
  \textit{et~al.}}{2008}]{2008A&A...481..827R}
\begin{barticle}
\bauthor{\bsnm{{Ruan}}, \binits{P.}},
\bauthor{\bsnm{{Wiegelmann}}, \binits{T.}},
\bauthor{\bsnm{{Inhester}}, \binits{B.}},
\bauthor{\bsnm{{Neukirch}}, \binits{T.}},
\bauthor{\bsnm{{Solanki}}, \binits{S.K.}},
\bauthor{\bsnm{{Feng}}, \binits{L.}}:
\byear{2008},
\batitle{{A first step in reconstructing the solar corona self-consistently
  with a magnetohydrostatic model during solar activity minimum}}.
\bjtitle{\aap}
\bvolume{481},
\bfpage{827}.
\doiurl{10.1051/0004-6361:20078834}.
\adsurl{2008A\%26A...481..827R}.
\end{barticle}
\endbibitem

\bibitem[\protect\citeauthoryear{{Sakurai}}{1981}]{1981SoPh...69..343S}
\begin{barticle}
\bauthor{\bsnm{{Sakurai}}, \binits{T.}}:
\byear{1981},
\batitle{{Calculation of force-free magnetic field with non-constant
  {$\alpha$}}}.
\bjtitle{\solphys}
\bvolume{69},
\bfpage{343}.
\doiurl{10.1007/BF00149999}.
\adsurl{1981SoPh...69..343S}.
\end{barticle}
\endbibitem

\bibitem[\protect\citeauthoryear{{Schl{\"u}ter} and
  {Temesv{\'a}ry}}{1958}]{1958IAUS....6..263S}
\begin{bchapter}
\bauthor{\bsnm{{Schl{\"u}ter}}, \binits{A.}},
\bauthor{\bsnm{{Temesv{\'a}ry}}, \binits{S.}}:
\byear{1958},
\bctitle{{The Internal Constitution of Sunspots}}.
In: \beditor{\bsnm{{Lehnert}}, \binits{B.}} (ed.)
\bbtitle{Electromagnetic Phenomena in Cosmical Physics},
\bsertitle{IAU Symp.}
\bseriesno{6},
\bfpage{263}.
\adsurl{1958IAUS....6..263S}.
\end{bchapter}
\endbibitem

\bibitem[\protect\citeauthoryear{{Socas-Navarro}}{2005}]{2005ApJ...631L.167S}
\begin{barticle}
\bauthor{\bsnm{{Socas-Navarro}}, \binits{H.}}:
\byear{2005},
\batitle{{The Three-dimensional Structure of a Sunspot Magnetic Field}}.
\bjtitle{\apjl}
\bvolume{631},
\bfpage{L167}.
\doiurl{10.1086/497334}.
\adsurl{2005ApJ...631L.167S}.
\end{barticle}
\endbibitem

\bibitem[\protect\citeauthoryear{{Spitzer}}{1958}]{4317418}
\begin{barticle}
\bauthor{\bsnm{{Spitzer}}, \binits{L.} \bsuffix{Jr.}}:
\byear{1958},
\batitle{{The Stellarator Concept}}.
\bjtitle{Physics of Fluids}
\bvolume{1},
\bfpage{253}.
\doiurl{10.1063/1.1705883}.
\adsurl{1958PhFl....1..253S}.
\end{barticle}
\endbibitem

\bibitem[\protect\citeauthoryear{{Steiner}, {Pneuman}, and
  {Stenflo}}{1986}]{1986A&A...170..126S}
\begin{barticle}
\bauthor{\bsnm{{Steiner}}, \binits{O.}},
\bauthor{\bsnm{{Pneuman}}, \binits{G.W.}},
\bauthor{\bsnm{{Stenflo}}, \binits{J.O.}}:
\byear{1986},
\batitle{{Numerical models for solar magnetic fluxtubes}}.
\bjtitle{\aap}
\bvolume{170},
\bfpage{126}.
\adsurl{1986A\%26A...170..126S}.
\end{barticle}
\endbibitem

\bibitem[\protect\citeauthoryear{{Thalmann}, {Wiegelmann}, and
  {Raouafi}}{2008}]{2008A&A...488L..71T}
\begin{barticle}
\bauthor{\bsnm{{Thalmann}}, \binits{J.K.}},
\bauthor{\bsnm{{Wiegelmann}}, \binits{T.}},
\bauthor{\bsnm{{Raouafi}}, \binits{N.-E.}}:
\byear{2008},
\batitle{{First nonlinear force-free field extrapolations of SOLIS/VSM data}}.
\bjtitle{\aap}
\bvolume{488},
\bfpage{L71}.
\doiurl{10.1051/0004-6361:200810235}.
\adsurl{2008A\%26A...488L..71T}.
\end{barticle}
\endbibitem

\bibitem[\protect\citeauthoryear{{Uchida} and
  {Low}}{1981}]{1981JApA....2..405U}
\begin{barticle}
\bauthor{\bsnm{{Uchida}}, \binits{Y.}},
\bauthor{\bsnm{{Low}}, \binits{B.C.}}:
\byear{1981},
\batitle{{Equilibrium configuration of the magnetosphere of a star loaded with
  accreted magnetized mass}}.
\bjtitle{J. Astrophys. Astron.}
\bvolume{2},
\bfpage{405}.
\doiurl{10.1007/BF02715550}.
\adsurl{1981JApA....2..405U}.
\end{barticle}
\endbibitem

\bibitem[\protect\citeauthoryear{{Valori}
  \textit{et~al.}}{2012}]{2012SoPh..278...73V}
\begin{barticle}
\bauthor{\bsnm{{Valori}}, \binits{G.}},
\bauthor{\bsnm{{Green}}, \binits{L.M.}},
\bauthor{\bsnm{{D{\'e}moulin}}, \binits{P.}},
\bauthor{\bsnm{{Vargas Dom{\'{\i}}nguez}}, \binits{S.}},
\bauthor{\bsnm{{van Driel-Gesztelyi}}, \binits{L.}},
\bauthor{\bsnm{{Wallace}}, \binits{A.}},
\bauthor{\bsnm{{Baker}}, \binits{D.}},
\bauthor{\bsnm{{Fuhrmann}}, \binits{M.}}:
\byear{2012},
\batitle{{Nonlinear Force-Free Extrapolation of Emerging Flux with a Global
  Twist and Serpentine Fine Structures}}.
\bjtitle{\solphys}
\bvolume{278},
\bfpage{73}.
\doiurl{10.1007/s11207-011-9865-8}.
\adsurl{2012SoPh..278...73V}.
\end{barticle}
\endbibitem

\bibitem[\protect\citeauthoryear{{Vernazza}, {Avrett}, and
  {Loeser}}{1981}]{1981ApJS...45..635V}
\begin{barticle}
\bauthor{\bsnm{{Vernazza}}, \binits{J.E.}},
\bauthor{\bsnm{{Avrett}}, \binits{E.H.}},
\bauthor{\bsnm{{Loeser}}, \binits{R.}}:
\byear{1981},
\batitle{{Structure of the solar chromosphere. III - Models of the EUV
  brightness components of the quiet-sun}}.
\bjtitle{\apjs}
\bvolume{45},
\bfpage{635}.
\doiurl{10.1086/190731}.
\adsurl{1981ApJS...45..635V}.
\end{barticle}
\endbibitem

\bibitem[\protect\citeauthoryear{{Wheatland}}{2004}]{2004SoPh..222..247W}
\begin{barticle}
\bauthor{\bsnm{{Wheatland}}, \binits{M.S.}}:
\byear{2004},
\batitle{{Parallel Construction of Nonlinear Force-Free Fields}}.
\bjtitle{\solphys}
\bvolume{222},
\bfpage{247}.
\doiurl{10.1023/B:SOLA.0000043579.93988.6f}.
\adsurl{2004SoPh..222..247W}.
\end{barticle}
\endbibitem

\bibitem[\protect\citeauthoryear{{Wheatland}}{2006}]{2006SoPh..238...29W}
\begin{barticle}
\bauthor{\bsnm{{Wheatland}}, \binits{M.S.}}:
\byear{2006},
\batitle{{A Fast Current-Field Iteration Method for Calculating Nonlinear
  Force-Free Fields}}.
\bjtitle{\solphys}
\bvolume{238},
\bfpage{29}.
\doiurl{10.1007/s11207-006-0232-0}.
\adsurl{2006SoPh..238...29W}.
\end{barticle}
\endbibitem

\bibitem[\protect\citeauthoryear{{Wheatland}}{2007}]{2007SoPh..245..251W}
\begin{barticle}
\bauthor{\bsnm{{Wheatland}}, \binits{M.S.}}:
\byear{2007},
\batitle{{Calculating and Testing Nonlinear Force-Free Fields}}.
\bjtitle{\solphys}
\bvolume{245},
\bfpage{251}.
\doiurl{10.1007/s11207-007-9054-y}.
\adsurl{2007SoPh..245..251W}.
\end{barticle}
\endbibitem

\bibitem[\protect\citeauthoryear{{Wheatland} and
  {R{\'e}gnier}}{2009}]{2009ApJ...700L..88W}
\begin{barticle}
\bauthor{\bsnm{{Wheatland}}, \binits{M.S.}},
\bauthor{\bsnm{{R{\'e}gnier}}, \binits{S.}}:
\byear{2009},
\batitle{{A Self-Consistent Nonlinear Force-Free Solution for a Solar Active
  Region Magnetic Field}}.
\bjtitle{\apjl}
\bvolume{700},
\bfpage{L88}.
\doiurl{10.1088/0004-637X/700/2/L88}.
\adsurl{2009ApJ...700L..88W}.
\end{barticle}
\endbibitem

\bibitem[\protect\citeauthoryear{{Wheatland}, {Sturrock}, and
  {Roumeliotis}}{2000}]{2000ApJ...540.1150W}
\begin{barticle}
\bauthor{\bsnm{{Wheatland}}, \binits{M.S.}},
\bauthor{\bsnm{{Sturrock}}, \binits{P.A.}},
\bauthor{\bsnm{{Roumeliotis}}, \binits{G.}}:
\byear{2000},
\batitle{{An Optimization Approach to Reconstructing Force-free Fields}}.
\bjtitle{\apj}
\bvolume{540},
\bfpage{1150}.
\doiurl{10.1086/309355}.
\adsurl{2000ApJ...540.1150W}.
\end{barticle}
\endbibitem

\bibitem[\protect\citeauthoryear{{Wiegelmann} and
  {Inhester}}{2003}]{2003SoPh..214..287W}
\begin{barticle}
\bauthor{\bsnm{{Wiegelmann}}, \binits{T.}},
\bauthor{\bsnm{{Inhester}}, \binits{B.}}:
\byear{2003},
\batitle{{Magnetic modeling and tomography: First steps towards a consistent
  reconstruction of the solar corona}}.
\bjtitle{\solphys}
\bvolume{214},
\bfpage{287}.
\doiurl{10.1023/A:1024282131117}.
\adsurl{2003SoPh..214..287W}.
\end{barticle}
\endbibitem

\bibitem[\protect\citeauthoryear{{Wiegelmann}
  \textit{et~al.}}{2007}]{2007A&A...475..701W}
\begin{barticle}
\bauthor{\bsnm{{Wiegelmann}}, \binits{T.}},
\bauthor{\bsnm{{Neukirch}}, \binits{T.}},
\bauthor{\bsnm{{Ruan}}, \binits{P.}},
\bauthor{\bsnm{{Inhester}}, \binits{B.}}:
\byear{2007},
\batitle{{Optimization approach for the computation of magnetohydrostatic
  coronal equilibria in spherical geometry}}.
\bjtitle{\aap}
\bvolume{475},
\bfpage{701}.
\doiurl{10.1051/0004-6361:20078244}.
\adsurl{2007A\%26A...475..701W}.
\end{barticle}
\endbibitem

\bibitem[\protect\citeauthoryear{{Wiegelmann}
  \textit{et~al.}}{2015}]{2015ApJ...815...10W}
\begin{barticle}
\bauthor{\bsnm{{Wiegelmann}}, \binits{T.}},
\bauthor{\bsnm{{Neukirch}}, \binits{T.}},
\bauthor{\bsnm{{Nickeler}}, \binits{D.H.}},
\bauthor{\bsnm{{Solanki}}, \binits{S.K.}},
\bauthor{\bsnm{{Mart{\'{\i}}nez Pillet}}, \binits{V.}},
\bauthor{\bsnm{{Borrero}}, \binits{J.M.}}:
\byear{2015},
\batitle{{Magneto-static Modeling of the Mixed Plasma Beta Solar Atmosphere
  Based on Sunrise/IMaX Data}}.
\bjtitle{\apj}
\bvolume{815},
\bfpage{10}.
\doiurl{10.1088/0004-637X/815/1/10}.
\adsurl{2015ApJ...815...10W}.
\end{barticle}
\endbibitem

\bibitem[\protect\citeauthoryear{{Zweibel} and
  {Hundhausen}}{1982}]{1982SoPh...76..261Z}
\begin{barticle}
\bauthor{\bsnm{{Zweibel}}, \binits{E.G.}},
\bauthor{\bsnm{{Hundhausen}}, \binits{A.J.}}:
\byear{1982},
\batitle{{Magnetostatic atmospheres - A family of isothermal solutions}}.
\bjtitle{Solar Physics}
\bvolume{76},
\bfpage{261}.
\doiurl{10.1007/BF00170987}.
\adsurl{1982SoPh...76..261Z}.
\end{barticle}
\endbibitem

\end{thebibliography}

\end{article} 
\end{document}